\documentclass[review]{elsarticle}
\usepackage[top=2.cm,bottom=2.cm,left=1.5cm,right=1.5cm]{geometry}
\usepackage{lineno,hyperref}
\usepackage{amssymb}
\usepackage{bm}
\usepackage{graphicx}
\usepackage{epstopdf, epsfig}
\usepackage{booktabs} 
\usepackage{array} 
\usepackage{amsmath}
\usepackage{mathtools}
\usepackage{graphicx,caption,subcaption,color} 
\usepackage{listings}
\usepackage{xcolor}
\usepackage{outlines}
\usepackage{CJKutf8}
\usepackage{float}
\usepackage{bigints}
\allowdisplaybreaks
\modulolinenumbers[5]
\usepackage[normalem]{ulem}

\journal{Journal of \LaTeX\ Templates}

\begin{document}

\begin{frontmatter}

\title{An effective correction method for droplet volume conservation in direct numerical simulation of droplet-laden turbulence}


\author[inst1]{Cheng Peng\corref{cor1}}
\affiliation[inst1]{Key Laboratory of High Efficiency and Clean Mechanical Manufacture, Ministry of Education, School of Mechanical Engineering, Shandong University, Jinan, 250061, Shandong, China}
\ead{pengcheng@sdu.edu.cn}
\cortext[cor1]{corresponding author}

\author[inst1]{Xuming Li}

\author[inst2]{Chunhua Zhang}
\affiliation[inst2]{School of Energy and Power Engineering, North University of China, Taiyuan 030051, Shanxi, China}

\author[inst3]{Lian-Ping Wang}%
\affiliation[inst3]{Guangdong Provincial Key Laboratory of Turbulence Research and Applications, Center for Complex Flows and Soft Matter Research and Department of Mechanics and Aerospace Engineering, Southern University of Science and Technology, Shenzhen 518055, Guangdong, China}

\author[inst4]{Xinnan Wu}%
\affiliation[inst4]{Shandong Scitech Innovation Group Co., Ltd., Jinan, 250101, Shandong, China}



\begin{abstract}
Accurately preserving the volume of the dispersed droplets remains a significant challenge in phase-field simulations of droplet-laden turbulence, especially under conditions that feature strong interfacial deformation and breakup. While modified phase-field equations have been developed to mitigate volume loss, their effectiveness has not been systematically assessed in the context of fully developed turbulent flows. In this work, we first evaluate the performance of several representative volume-corrected phase-field models in direct numerical simulations of droplet-laden homogeneous isotropic turbulence. Our results reveal that, at sufficiently high Weber numbers, none of the existing models provides satisfactory droplet-volume preservation. To address this limitation, we then propose a simple yet effective modification of the conservative Allen–Cahn equation by incorporating a curvature-dependent counter-diffusion correction. Direct numerical simulations in turbulent regimes demonstrate that the proposed model achieves conservation of droplet volume in a statistical sense, while avoiding common adverse effects, such as numerical instability, violation of global mass conservation, increased computational cost, artificial coarsening, or enhanced spurious velocities.
\end{abstract}


\end{frontmatter}

\section{Introduction}\label{sec:introduction}

Direct numerical simulations (DNS) of turbulent flows laden with dispersed droplets or bubbles (hereafter referred to as droplet-laden turbulence for brevity) are critical for advancing the understanding of a wide range of natural phenomena and engineering applications. These include breaking ocean waves, immiscible groundwater mixing, fuel injection, emulsion polymerization, and the operation of bubble column reactors. 

Among contemporary numerical approaches, the phase-field method has become a widely adopted tool for DNS of droplet-laden turbulence~\cite{komrakova2015numerical,scarbolo2016turbulence,alessio2017viscosity,soligo2019breakage,lai2022systematic}. Despite its extensive use, a fundamental and persistent challenge remains: the inadequate preservation of the volume of the dispersed phase during simulations. This well-documented limitation seriously hinders the broader application and quantitative reliability of phase-field methods in the study of turbulent multiphase flows~\cite{elghobashi2019direct}.

The phase-field method is a diffuse-interface approach that implicitly represents fluid interfaces through the evolution of an order parameter, governed by equations, such as the Cahn-Hilliard (CH) equation or the Allen-Cahn (AC) equation. Owing to their diffuse-interface property, phase-field methods are particularly well suited for simulating complex interfacial dynamics involving large deformations, breakup, and coalescence, without the need for explicit interface tracking. 

The phase-field formulation drives the evolution of a system by the minimizing a prescribed free-energy functional. Consequently, when a droplet is immersed in a surrounding fluid, its size may decrease spontaneously in order to reduce the total free energy, a phenomenon commonly referred to as spontaneous droplet shrinkage~\cite{yue2007spontaneous}. From a thermodynamic standpoint, such shrinkage is physically admissible. However, in practical simulations, phase-field methods employ diffuse interfaces with a finite thickness spanning several grid points, which is typically orders of magnitude larger than the physical interface thickness in real flows of hydrodynamic interest. This disparity in length scales leads to a significant numerical amplification of the shrinkage process, causing droplet volume to diminish over drastically shorter time scales than would occur physically.

In static or weakly disturbed flows, droplet shrinkage can be partially mitigated if the initial free energy exceeds a certain threshold~\cite{yue2007spontaneous}. In turbulent flows, however, strong shear and continuous deformation constantly alter the droplet morphology, rendering this threshold increasingly restrictive~\cite{wagner1997effect}. The situation is further exacerbated when droplets undergo repeated breakup and coalescence, as this process generates small satellite droplets that are especially susceptible to rapid dissolution due to local free-energy minimization.

To mitigate droplet volume loss, a variety of correction strategies have been proposed~\cite{huang2014mass,wang2015mass,li2016phase,zhang2017flux,zhang2019interface,bao2024phase,yang2025two}. 
Among the earliest approaches are those that enforce global conservation of the dispersed-phase mass. For instance, Huang et al.~\cite{huang2014mass} proposed such a correction for an axisymmetric phase-field lattice Boltzmann formulation, in which an additional time-derivative term is introduced and applied across the entire domain. This term is proportional to the deviation of the total droplet volume from its initial value and scaled by an empirically determined coefficient.

Subsequently, Wang et al.~\cite{wang2015mass} proposed a global mass-correction strategy that eliminates the need for an empirical coefficient. In their model, the deviation of the total droplet mass from its initial value is evaluated at each time step and compensated by a spatially uniform correction applied only within the interfacial region. While designed to enforce strict global volume conservation, this approach suffers from several inherent limitations.  First, droplet volume loss originates from localized dissolution processes, whereas the total mass of the order parameter is already conserved; introducing additional mass therefore risks violating global mass conservation unless further corrective measures are applied. Second, droplets of different sizes typically dissolve at different rates, with smaller droplets being more susceptible to dissolution. Uniform redistribution of the dissolved mass thus lacks a clear physical basis and may induce artificial coarsening, characterized by unphysical mass transfer from small droplets to larger ones. Third, in strongly turbulent flows involving pronounced deformation and frequent topological changes, accurately identifying and tracking droplet volume becomes increasingly difficult, especially with diffuse interfaces. Errors in volume estimation may, paradoxically, lead to overcompensation and further degradation of droplet volume conservation. A detailed discussion of this issue is provided in Appendix~\ref{sec:appendixB}.

A more widely adopted strategy involves introducing localized correction terms directly into the phase-field equations. When formulated in divergence form, these corrections ensure that mass redistribution remains local while preserving global mass conservation. A representative model in this category is the conservative Allen–Cahn (CAC) equation~\cite{chiu2011conservative}, which suppresses diffusion of the order parameter normal to the interface to prevent unphysical mass transfer across the interface. Our recent numerical tests indicate that the CAC equation can preserve droplet volume even when the droplet size approaches the interface thickness, provided that droplet deformation remains weak. However, under conditions of strong deformation or breakup, significant droplet volume loss is still observed~\cite{li2025comparativestudycriticalassessment}.

Beyond the AC-based model,
Li et al. proposed a profile-corrected CH (CH-PC) model~\cite{li2016phase}, which introduces a weighted diffusion term similar to the CAC formulation to mitigate interface dispersion. Here, interface dispersion refers to the progressive thickening of the interface and the deviation of the order parameter beyond its theoretical bounds. As the weighting parameter increases, the CH-PC model asymptotically approaches the behavior of the CAC equation. Building on this framework, Zhang and Ye proposed a flux-corrected variant (CH-FC) that explicitly subtracts the normal component of the CH flux~\cite{zhang2017flux}. Although these models can reduce numerical diffusion and partially alleviate droplet volume loss, they do not fully resolve the problem. For example, Soligo et al.~\cite{soligo2019mass} reported that in droplet-laden turbulent channel flow simulations—characterized by relatively weak droplet deformation and a small Cahn number of 0.02, which is known to suppress volume loss~\cite{komrakova2015numerical}—the CH-FC and CH-PC models still exhibited droplet volume losses of approximately 0.5\% and 2.5\%, respectively, over relatively short simulation times.

An alternative localized correction was proposed by Zhang et al.~\cite{zhang2019interface}, which may be interpreted as a velocity-weighted extension of the CAC correction. Termed the interface-compressed CH (CH-IC) model, it was shown to effectively suppress interface dispersion in simple two-dimensional configurations, such as Rayleigh–Taylor instability and droplet stretching in laminar shear flows. However, its effectiveness in suppressing droplet volume loss in fully developed droplet-laden turbulence has not yet been systematically examined.

More recently, a third line of research has emerged, focusing on modifying the mobility in phase-field equations to suppress unphysical diffusion. Bao and Guo proposed an order-parameter-dependent mobility that maintains nominal diffusivity near the interface while vanishing in the bulk phases~\cite{bao2024phase}. Our own tests indicate that this approach does not fully resolve droplet volume loss in droplet-laden turbulent simulations~\cite{li2025comparativestudycriticalassessment}. Yang and Kim proposed an alternative formulation in which the mobility decreases with increasing local curvature, thereby suppressing the dissolution of small droplets characterized by large curvature~\cite{yang2025two}. Although this model was tested in complex three-dimensional configurations, such as the dam-break problem, its effectiveness in droplet-laden turbulence remains an open question, as confirmed through personal communication with the authors.

In the present work, we propose a simple yet effective approach to conserve droplet volume in direct numerical simulations of droplet-laden turbulence, addressing a persistent limitation of phase-field methods in this context. By incorporating a curvature-dependent counter-diffusive correction into the CAC equation, we enforce droplet volume conservation in \emph{a statistical sense}. This correction may be interpreted as a precipitation-like mechanism that counteracts the continuous dissolution of small droplets driven by free-energy minimization. The effectiveness of the proposed method is demonstrated through three-dimensional simulations of droplet-laden homogeneous isotropic turbulence over a range of Weber numbers. The results confirm that the proposed approach achieves statistical droplet volume conservation without inducing artificial coarsening, violating global mass conservation, or degrading numerical stability or computational efficiency.

The remainder of this paper is organized as follows. Section~\ref{sec:models} reviews the phase-field formulation and existing correction strategies for droplet volume conservation. Section~\ref{sec:previous} assesses and compares their performances in droplet-laden turbulence. Section~\ref{sec:modifiedCAC} introduces the proposed model in detail. Numerical validations and comparative analyses are presented in Section~\ref{sec:validation}. Finally, Section~\ref{sec:conclusion} summarizes the main findings. Additional details of the lattice Boltzmann solver and an estimation of droplet volume variation in phase-field simulations are provided in Appendices~\ref{sec:appendixA} and~\ref{sec:appendixB}, respectively.

\section{Phase-field equations and previous efforts to conserve droplet volume}\label{sec:models}

Phase-field methods describe multiphase flows by evolving an order parameter, $\phi$, governed by a convection–diffusion-type equation. The order parameter serves as a diffused representation of the interface, across which material properties such as density and viscosity are smoothly interpolated between their respective bulk-phase values. Following the notation adopted for droplet-laden turbulence, the order parameter takes the values $\phi_{\rm D}$ and $\phi_{\rm C}$ in the dispersed (droplet) phase and the carrier phase, respectively. The interface is commonly identified with the iso-surface $\phi_0 = (\phi_{\rm D} + \phi_{\rm C})/2$.

Among various phase-field formulations, the CH equation is one of the most widely used models for incompressible multiphase flows. It reads~\cite{kendon2001inertial,badalassi2003computation,jacqmin1999calculation}:
\begin{equation}
\frac{\partial \phi}{\partial t} + \boldsymbol{\nabla} \cdot (\phi \bm{u})
= \boldsymbol{\nabla} \cdot \big( M_{\rm CH} \boldsymbol{\nabla} \mu_\phi \big),
\label{eq:CH}
\end{equation}
where $\bm{u}$ is the fluid velocity, $M_{\rm CH}$ is the mobility coefficient controlling the diffusivity of $\phi$, and $\mu_\phi$ is the chemical potential. Considering the Ginzburg-Landau free energy and a double-well potential, the following chemical potential can be obtained,
\begin{equation}
\mu_{\phi} = 4\beta (\phi-\phi_{\rm D})(\phi-\phi_{\rm C})(\phi-\phi_{0})
- \kappa \nabla^{2}\phi.
\label{eq:chemical_potential}
\end{equation}
The coefficients $\beta$ and $\kappa$ are related to the surface tension $\sigma$ and the diffuse-interface thickness $W$ through 
\begin{equation}
\beta = \frac{12\sigma}{W |\phi_{\rm D} - \phi_{\rm C}|^4}, \qquad \kappa = \frac{3\sigma W}{2|\phi_{\rm D} - \phi_{\rm C}|^2}. 
\label{eq:coefficient1}
\end{equation}

As discussed in Section~\ref{sec:introduction}, numerical solutions of the CH equation may suffer from interface smearing and an apparent loss of droplet volume, particularly in turbulent flows where strong interfacial deformation and large curvature variations are present. To alleviate these deficiencies, numerous modified CH formulations have been proposed. Many of them can be written in a unified form
\begin{equation}
  \frac{\partial \phi}{\partial t} + \boldsymbol{\nabla} \cdot (\phi \bm{u})
= \boldsymbol{\nabla} \cdot \big( M_{\rm CH} \boldsymbol{\nabla} \mu_\phi \big) + Q,
\label{eq:modifiedCH}
\end{equation}
where $Q$ denotes an additional correction term introduced to counteract interface diffusion or mass loss.

In Wang et al.~\cite{wang2015mass}'s global mass-correction model, the correction term is explicitly determined from the deviation of the total dispersed-phase mass:
\begin{equation}
    Q = Q_{\rm G} = \left\{\begin{array}{cc}
         &\dfrac{1}{\Delta V}\left[\dfrac{M(t) - M(0)}{\delta t \Delta\rho} - \bigintss_{\Omega_{\rm C}}\boldsymbol{\nabla}\cdot\left(M_{\rm CH}\nabla\mu_\phi\right)dV\right],\qquad \phi\in[\phi_1,\phi_2],\\
         & 0, \qquad \qquad\qquad\qquad\qquad\qquad\qquad\qquad\qquad \qquad \qquad{\rm elsewhere}
    \end{array} \right..
    \label{eq:wang2015}
\end{equation}
Here, the subscript $\rm G$ emphasizes the global nature of this correction. The quantity $\Delta V$ denotes the volume of the interfacial region bounded by $\phi_1$ and $\phi_2$; $M(t)$ and $M(0)$ are the dispersed-phase masses at the current and initial times, respectively; $\delta t$ is the time-step size; and $\Delta\rho = \rho_{\rm D} - \rho_{\rm C}$ is the density difference between the two phases. A commonly adopted choice is $\phi_1 = \phi_{\rm C} + 0.1(\phi_{\rm D} - \phi_{\rm C})$ and $\phi_2 = \phi_{\rm C} + 0.9(\phi_{\rm D} - \phi_{\rm C})$. The region $\Omega_{\rm C}$ represents the volume occupied by the carrier phase. In this formulation, the deviation of the total droplet mass from its initial value is evaluated at each time step, and a uniform compensating source term is applied within the interfacial region to restore the prescribed global mass.

In contrast to the global correction, Li et al.~\cite{li2016phase} proposed a localized profile-corrected term,
\begin{equation}
   Q = Q_{\rm CH-PC} = \boldsymbol{\nabla}\cdot\left\{\lambda_{\rm PC}M_{\rm CH}\left[\boldsymbol{\nabla}\phi + \frac{4}{W}\frac{\left(\phi - \phi_{\rm D}\right)\left(\phi - \phi_{\rm C}\right)}{\phi_{\rm D} - \phi_{\rm C}}{\bm n}\right]\right\},
   \label{eq:Li2016}
\end{equation}
where ${\bm n} = \boldsymbol{\nabla \phi}/\vert\boldsymbol{\nabla \phi}\vert$ denotes the interface normal, and $\lambda_{\rm PC}$ is a tunable parameter controlling the strength of the correction. Increasing $\lambda_{\rm PC}$ enhances the effectiveness of the correction, but may reduce numerical stability.

Building on this approach, Zhang and Ye~\cite{zhang2017flux} introduced a flux-corrected term,
\begin{equation}
   Q = Q_{\rm CH-FC} = \boldsymbol{\nabla}\cdot\left\{\lambda_{\rm PC}M_{\rm CH}\left[\boldsymbol{\nabla}\phi + \frac{4}{W}\frac{\left(\phi - \phi_{\rm D}\right)\left(\phi - \phi_{\rm C}\right)}{\phi_{\rm D} - \phi_{\rm C}}{\bm n}\right] - M_{\rm CH}\left({\bm n}\cdot\boldsymbol{\nabla}\mu_\phi\right){\bm n}\right\},
   \label{eq:Zhang2019a}
\end{equation}
which explicitly removes the diffusive flux normal to the interface in the original CH equation, thereby enhancing both mass conservation and interface sharpness.

Alternatively, Zhang et al.~\cite{zhang2019interface} proposed an interface-compressed correction,
\begin{equation}
   Q = Q_{\rm CH-IC} = \boldsymbol{\nabla}\cdot\left[\lambda_{\rm IC}\vert \boldsymbol{\nabla}\left({\bm u}\cdot{\bm n}\right)\vert\left(\phi - \phi_{\rm D}\right)\left(\phi - \phi_{\rm C}\right){\bm n}\right],
   \label{eq:Zhang2019b}
\end{equation}
where $\lambda_{\rm IC}$ is a strength parameter that must be chosen to balance the effectiveness of the correction with numerical stability.

Beyond additive correction terms, another strategy is to modify the mobility itself and locally modify the interface diffusion flux. Bao and Guo~\cite{bao2024phase} introduced a singular mobility formulation,
\begin{equation}
    M_{\rm CH-B} = \frac{4}{\left(\phi_{\rm D} -\phi_{\rm L}\right)^2}M_{\rm CH}\vert \phi - \phi_{\rm D}\vert \vert \phi - \phi_{\rm L}\vert,
    \label{eq:bao2024}
\end{equation}
in which the local mobility $M_{\rm CH-B}$ smoothly vanishes in the bulk phases as $\phi$ approaches either $\phi_{\rm D}$ or $\phi_{\rm C}$, while recovering the nominal mobility $M_{\rm CH}$ at the interface $\phi = \phi_0$.

Alternatively, Yang and Kim~\cite{yang2025two} proposed a curvature-dependent singular mobility,
\begin{equation}
    M_{\rm CH-Y} = \frac{\left(\phi - \phi_0\right)^2}{1 + \gamma\vert \boldsymbol{\nabla} \cdot {\bm n}\vert },
    \label{eq:yang2025}
\end{equation}
where $\gamma > 0$ is a large scaling parameter that reduces mobility in regions of high curvature, thereby mitigating the dissolution of small droplets. 

All the above corrections are based on the CH equation. Alternatively, the AC equation is also widely used in multiphase flow simulations~\cite{allen1979microscopic}. Since the original AC equation does not conserve mass, its conservative variant is more commonly employed. In particular, the CAC equation introduces a local counter term associated with interface curvature~\cite{chiu2011conservative},
\begin{equation}
    \frac{\partial \phi}{\partial t} + \boldsymbol{\nabla} \cdot (\phi \bm{u}) 
=  \boldsymbol{\nabla} \cdot \left\{M_{\rm AC}\left[ \boldsymbol{\nabla} \phi + \frac{4}{W} \frac{(\phi - \phi_{\rm D})(\phi - \phi_{\rm C})}{\phi_{\rm D} - \phi_{\rm C}}{\bm n} \right]\right\},
\end{equation}
where $M_{\rm AC}$ denotes the mobility in AC-type formulations. Although both $M_{\rm CH}$ and $M_{\rm AC}$ are referred to as mobilities, they have different physical dimensions. Unlike material transport coefficients such as viscosity or thermal conductivity, mobility is not a material property with a well-defined physical value in the hydrodynamic limit. Rather, it is treated as a numerical parameter, chosen to balance accuracy and stability.

\section{Performance of previous models on droplet volume conservation in droplet-laden turbulence }\label{sec:previous}
While the aforementioned volume-preserving models have been shown to be effective in laminar flow tests, most of them have not been evaluated in fully developed droplet-laden turbulence at large Weber numbers. Consequently, their ability to address the long-standing problem of droplet volume loss in phase-field simulations of droplet-laden turbulence remains unclear.

To systematically assess the capabilities of existing phase-field models in preserving droplet volume under turbulent conditions, we extend our previous comparative study~\cite{li2025comparativestudycriticalassessment} to all models introduced in Section~\ref{sec:models}. These include the original CH equation~\cite{jacqmin1999calculation}, the CAC formulation~\cite{chiu2011conservative}, various modified CH models (CH-PC, CH-FC, CH-IC)~\cite{li2016phase,zhang2017flux,zhang2019interface}, singular-mobility approaches~\cite{bao2024phase,yang2025two} (CH-B, CH-Y), and the global mass-correction method~\cite{wang2015mass} (CH-G).

\begin{figure}
    \centering
    \includegraphics[width=0.5\textwidth]{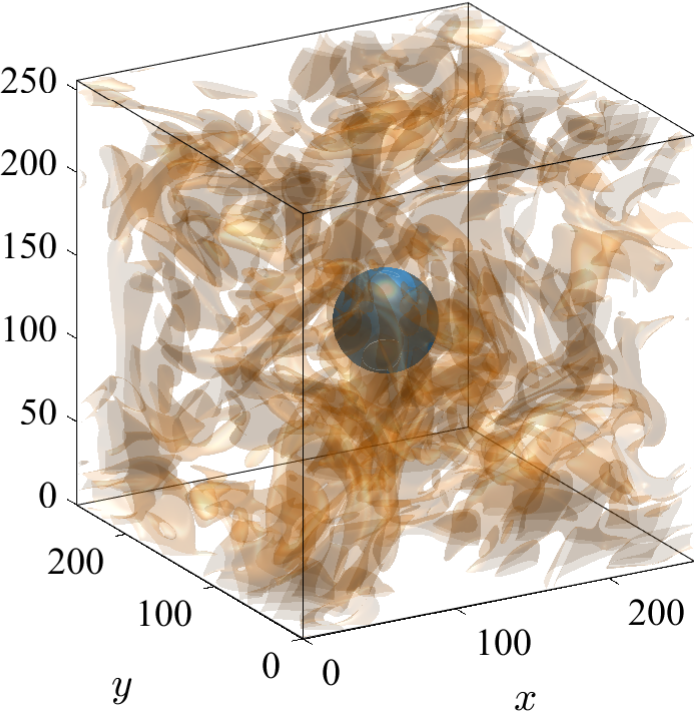}
    \caption{Schematic of a droplet in homogeneous isotropic turbulence. }
    \label{fig:schematic}
\end{figure}

The test problem consists of a single large droplet embedded in fully developed homogeneous isotropic turbulence (HIT), generated using the stochastic forcing scheme of Eswaran and Pope~\cite{eswaran1988examination}, with a Taylor-microscale Reynolds number $Re_\lambda = 42.28$ (Figure~\ref{fig:schematic}). The droplet is initially placed at the center of a cubic domain of size $L = 256 \delta x$ ($\delta x$ is the grid spacing) and initialized using a three-dimensional hyperbolic tangent profile:
\begin{equation}
\phi(x,y,z) = \frac{\phi_{\rm D} + \phi_{\rm C}}{2} 
+ \frac{\phi_{\rm D} - \phi_{\rm C}}{2} \, 
\tanh \left\{\frac{2 \Big[ \sqrt{(x - 0.5 L)^2 + (y - 0.5 L)^2 + (z - 0.5 L)^2} - r_0 \Big]}{W}\right\},
\label{eq:init-phi-droplet-3D}
\end{equation}
where $r_0 = 30 \delta x$ is the initial droplet radius. Periodic boundary conditions are applied in all directions. The stochastic forcing is applied continuously after droplet insertion to maintain turbulence.

The simulations are carried out using the lattice Boltzmann (LB) method, which serves as a second-order accurate, indirect solver for both the phase-field and Navier–Stokes equations. Since droplet volume conservation is primarily dictated by the phase-field formulation rather than the solver itself, a detailed description of the phase-field LB implementation is provided in Appendix~\ref{sec:appendixA}. Full validation of the LB solver is omitted here; interested readers are referred to our previous work for comprehensive tests and comparisons of representative phase-field LB models across multiple benchmark problems~\cite{li2025comparativestudycriticalassessment}.

We first examine five phase-field models—CAC, CH, CH-PC, CH-FC, and CH-IC—at a Weber number ${\rm We}_{\rm d} = 20.0$, defined as ${\rm We}_{\rm d} = \rho_{\rm D} u'^2 (2 r_0)/\sigma$, where $\rho_{\rm D}$ is the dispersed-phase density, $u'$ is the turbulence root-mean-square velocity, and $r_0$ is the initial droplet radius. Both phases are assigned equal densities, $\rho_D = \rho_C = 1.0$, and the interface thickness is set to $W = 6 \delta x$. The choice of equal densities and relatively thick interface is motivated by the need to maintain numerical stability in CH-based models under strong droplet deformation, stretching, and breakage. The LB solver’s low numerical dissipation further constrains stability~\cite{dellar2001bulk,barad2017lattice,wissocq2022hydrodynamic}.
The kinematic viscosities of both phases are set to $\nu_{\rm D} = \nu_{\rm C} = 0.006 \delta x^2/\delta t$, and the constant mobility is $0.1$ for both $M_{\rm CH}$ in CH-type models and $M_{\rm AC}$ in AC-type models, despite their differing physical dimensions. The order-parameter limits $\phi_D = 1$ and $\phi_C = 0$ are used consistently across all simulations.

Under these parameters, models with singular mobilities (CH-B and CH-Y) are excluded due to numerical instability. The global mass-correction model CH-G is also omitted, as the equal-density setup yields $\Delta \rho = 0$ in Eq.~(\ref{eq:wang2015}). For comparison, results from our proposed model (introduced in Section~\ref{sec:modifiedCAC}) are included.

\begin{figure}
    \centering
    \includegraphics[width=0.5\textwidth]{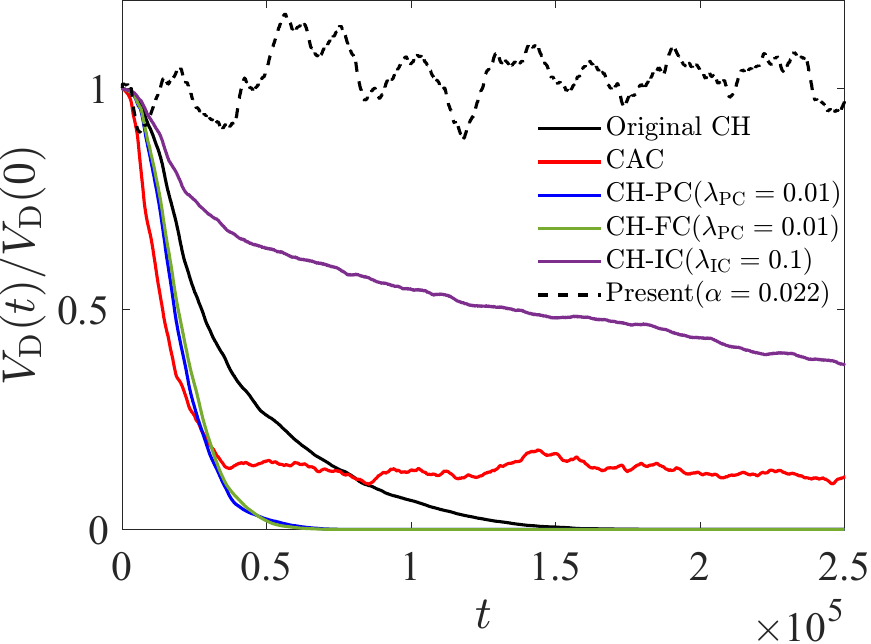}
    \caption{The evolution of droplet volume for CAC, CH, CH-PC, CH-FC, CH-IC, and our modified CAC models}
    \label{fig:WE20_DropletVolume_6Models}
\end{figure}

Figure~\ref{fig:WE20_DropletVolume_6Models} shows the temporal evolution of droplet volume for the six models. Tunable parameters are set as $\lambda_{\rm PC} = 0.01$ for CH-PC and CH-FC, and $\lambda_{\rm IC} = 0.1$ for CH-IC. All previously proposed models exhibit substantial droplet volume loss. CH-PC and CH-FC, in particular, lose volume faster than the original CH model. This behavior arises from the interplay of two effects. The relatively thick interface ($W = 6 \delta x$) causes CAC-like corrections in CH-PC and CH-FC to accelerate volume loss at early times. Later, while CAC itself maintains residual droplets due to the reduction of the effective Weber number after fragmentation, CH-type models continue to suffer spontaneous shrinkage, whereby small droplets dissolve continuously~\cite{yue2007spontaneous}. Consequently, CH-PC and CH-FC combine the initial CAC-like loss with persistent CH-type shrinkage, resulting in the poorest overall performance. It is worth noting that with a thinner interface (if stability permits), CH-PC and CH-FC can substantially improve droplet volume preservation compared to the original CH model, consistent with previous reports~\cite{soligo2019mass}.

Our proposed model, by comparison, statistically preserves the droplet volume near its initial value. While instantaneous deviations exceeding 15\% are observed, these are expected within the framework of phase-field simulations with diffuse interfaces (see Appendix~\ref{sec:appendixB}). In fact, strictly enforcing the droplet volume can be overly restrictive and unphysical. 

\begin{figure}[htbp]
    \centering
    \begin{subfigure}[b]{0.32\textwidth}
        \centering
        \includegraphics[width=\textwidth]{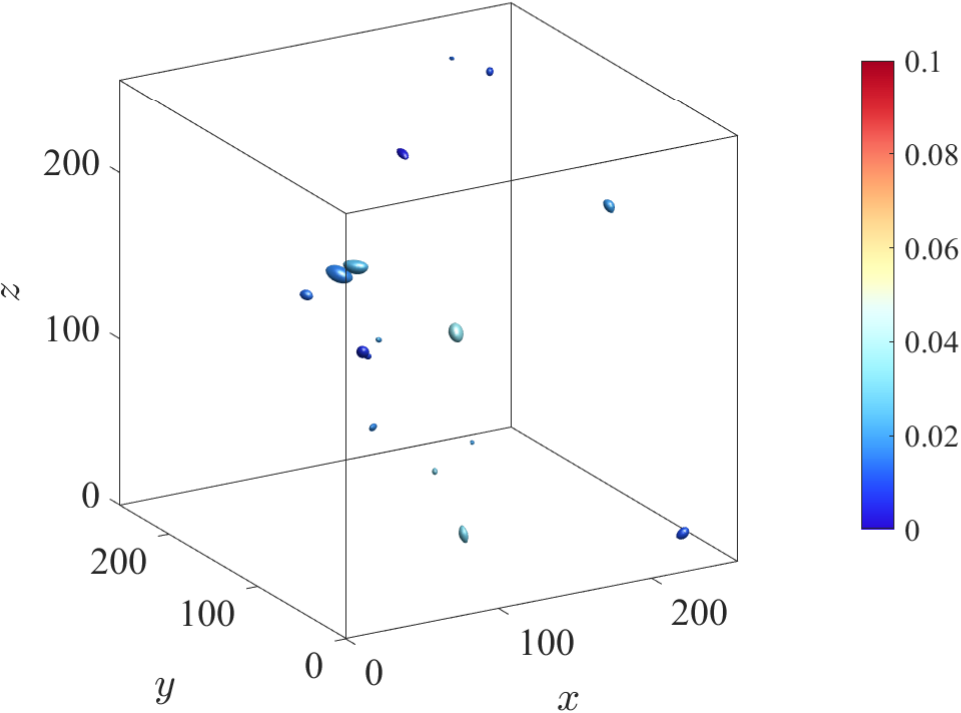}
        \caption{CH-PC}
        \label{fig:WE20_cloud_CHPC}
    \end{subfigure}
    \hfill
    \begin{subfigure}[b]{0.32\textwidth}
        \centering
        \includegraphics[width=\textwidth]{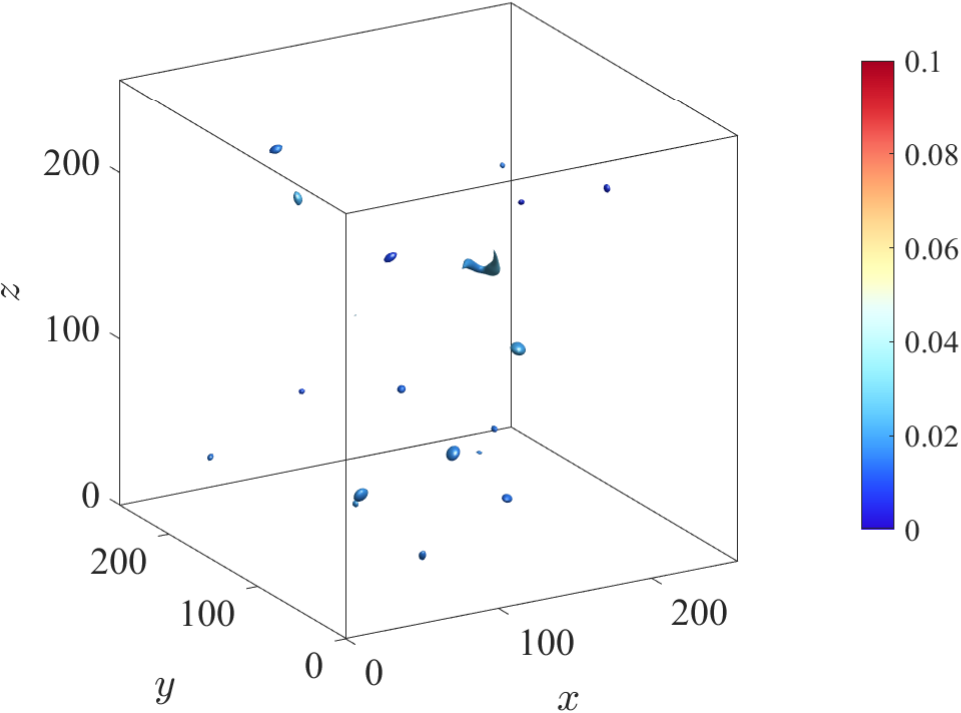}
        \caption{CH-FC}
        \label{fig:WE20_cloud_CHFC}
    \end{subfigure}
    \hfill
    \begin{subfigure}[b]{0.32\textwidth}
        \centering
        \includegraphics[width=\textwidth]{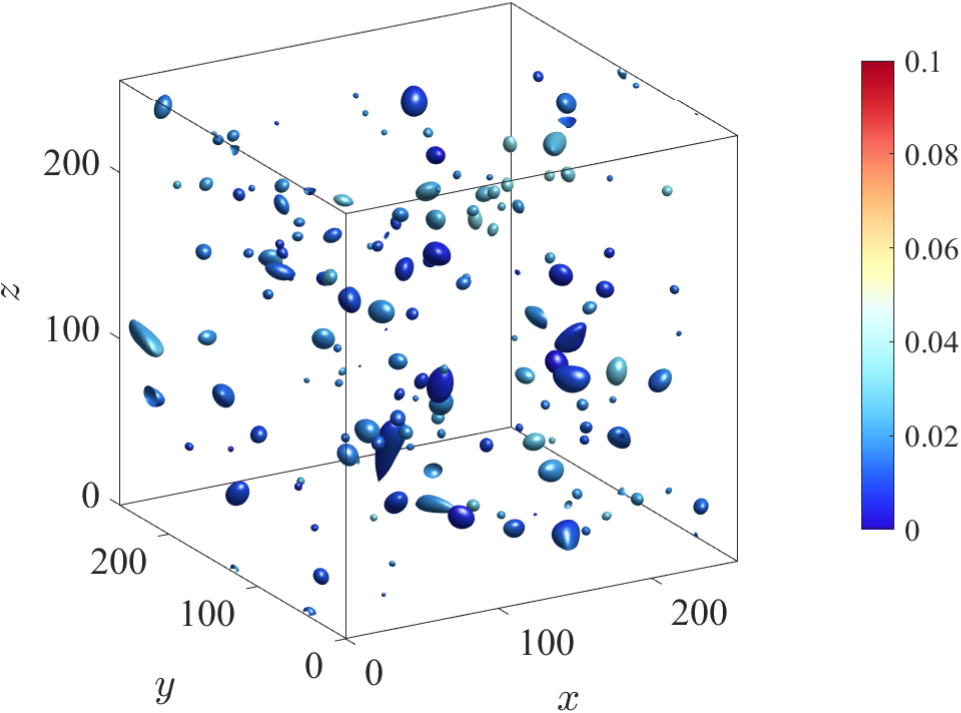}
        \caption{CH-IC}
        \label{fig:WE20_cloud_CHIC}
    \end{subfigure}

    \vspace{1em} 

    \begin{subfigure}[b]{0.32\textwidth}
        \centering
        \includegraphics[width=\textwidth]{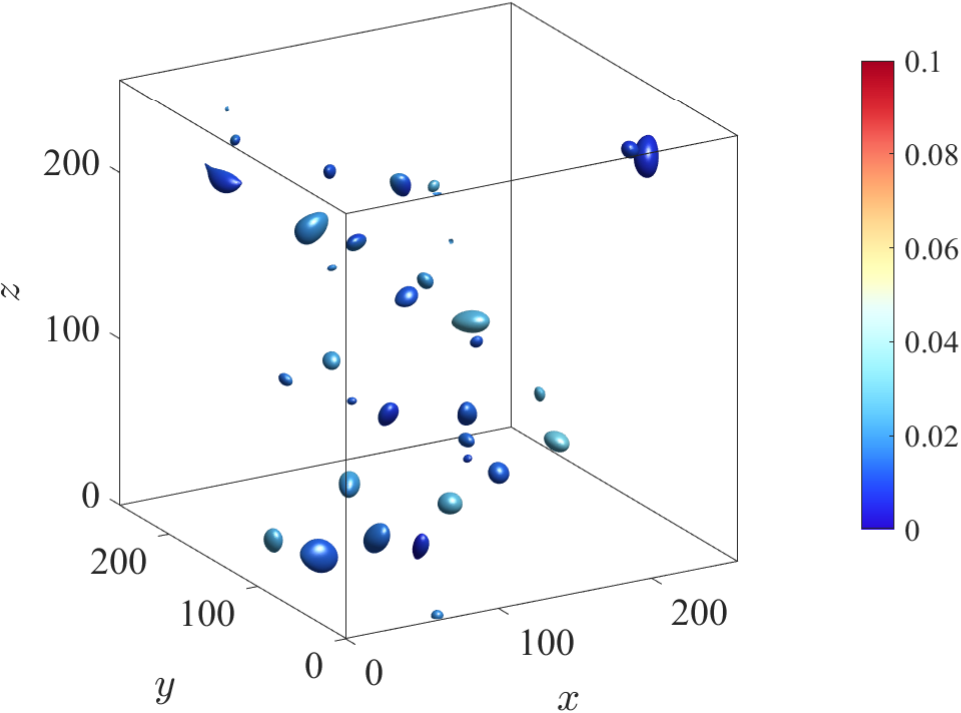}
        \caption{CH}
        \label{fig:WE20_cloud_CH}
    \end{subfigure}
    \hfill
    \begin{subfigure}[b]{0.32\textwidth}
        \centering
        \includegraphics[width=\textwidth]{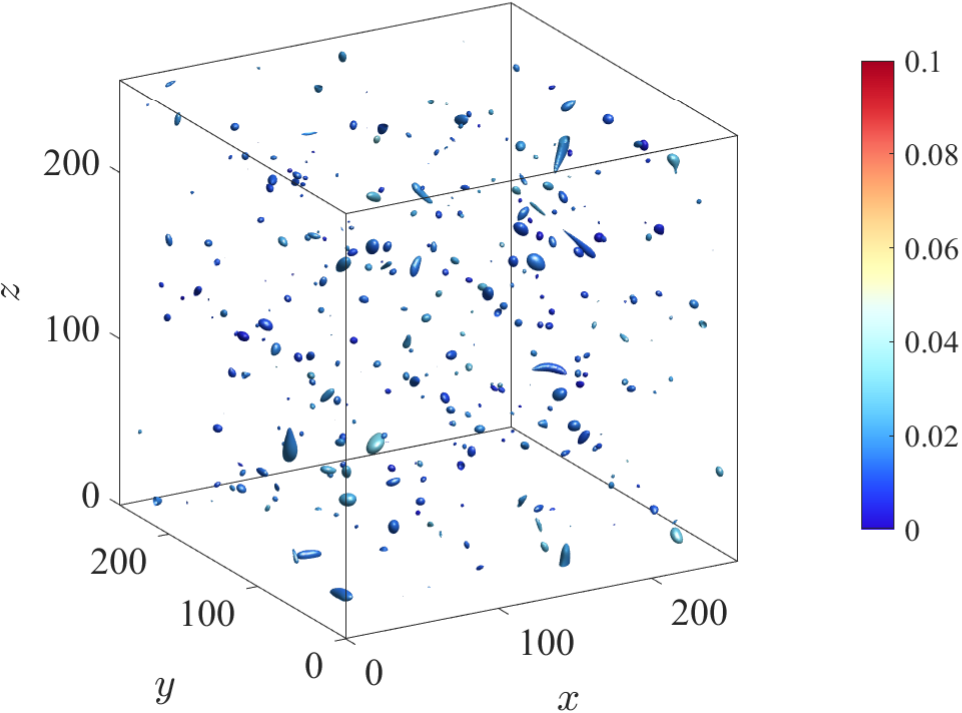}
        \caption{CAC}
        \label{fig:WE20_cloud_CAC}
    \end{subfigure}
    \hfill
    \begin{subfigure}[b]{0.32\textwidth}
        \centering
        \includegraphics[width=\textwidth]{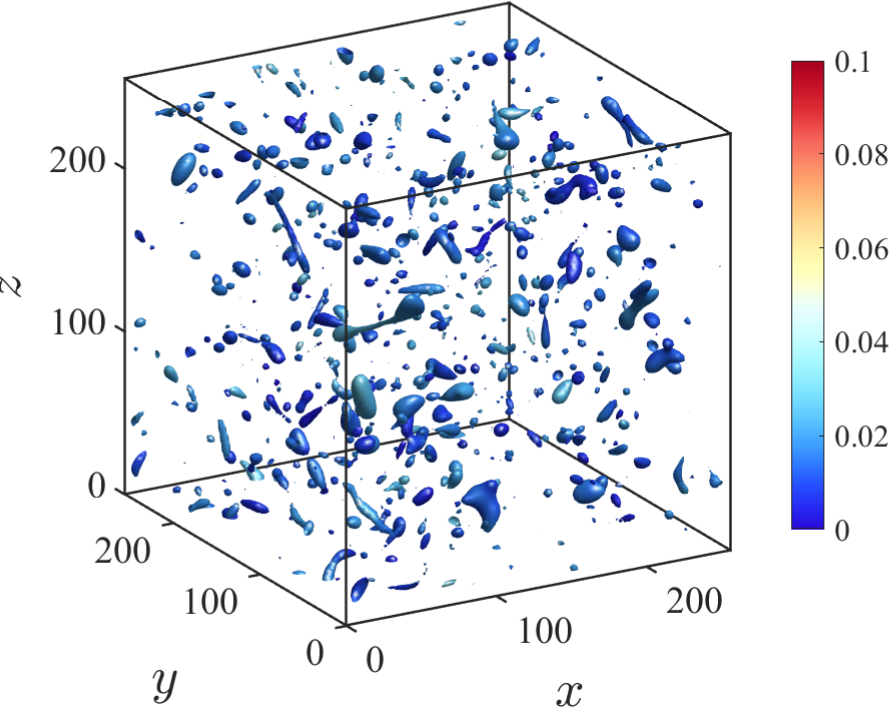}
        \caption{Present}
        \label{fig:WE20_cloud_Present}
    \end{subfigure}

    \caption{Visualizations of the two-phase interface at time step $=50,000$ at ${\rm We}_{\rm d} = 20$. The colorbar in each subplot shows the velocity magnitude in the lattice unit.}
    \label{fig:WE20_cloud_6models}
\end{figure}

Figure~\ref{fig:WE20_cloud_6models} presents snapshots of the two-phase interfaces at time step $t = 50,000$. Among the five previous models, only CH-IC retains a significant fraction of the droplet volume. In the original CH model, most small droplets are lost due to spontaneous shrinkage, while only larger droplets remain. The CAC model preserves many small droplets but still suffers considerable volume loss overall. The CH-PC and CH-FC models nearly lose all droplets, reflecting their combined early-stage CAC-like loss and CH-type shrinkage. In contrast, our proposed model not only preserves the overall droplet volume but also maintains both large and small droplets effectively. 

For the three modified CH models—CH-PC, CH-FC, and CH-IC—the respective tunable parameters, $\lambda_{\rm PC}$ in CH-PC and CH-FC, and $\lambda_{\rm IC}$ in CH-IC, play a significant role in numerical performance. To assess this influence, we conducted simulations with varying parameter values. For CH-PC and CH-FC, $\lambda_{\rm PC}$ was varied over $[0.01, 0.5]$, and we observed that the rate of droplet volume loss increases with $\lambda_{\rm PC}$. Consequently, a smaller correction term yields better droplet volume preservation. 
This seemingly counterintuitive trend is related to the relatively large interface thickness ($W = 6 \delta x$) adopted for numerical stability. 
As noted in our recent comparative study~\cite{li2025comparativestudycriticalassessment}, a wider interface favors droplet volume conservation in CH-type LB models, whereas it tends to reduce conservation in AC-type LB models. Since the correction term in CH-PC (and, by inheritance, CH-FC) essentially resembles the CAC-type term, its poorer volume-conservation performance under these conditions is understood.

\begin{figure}
    \centering
    \includegraphics[width=0.5\textwidth]{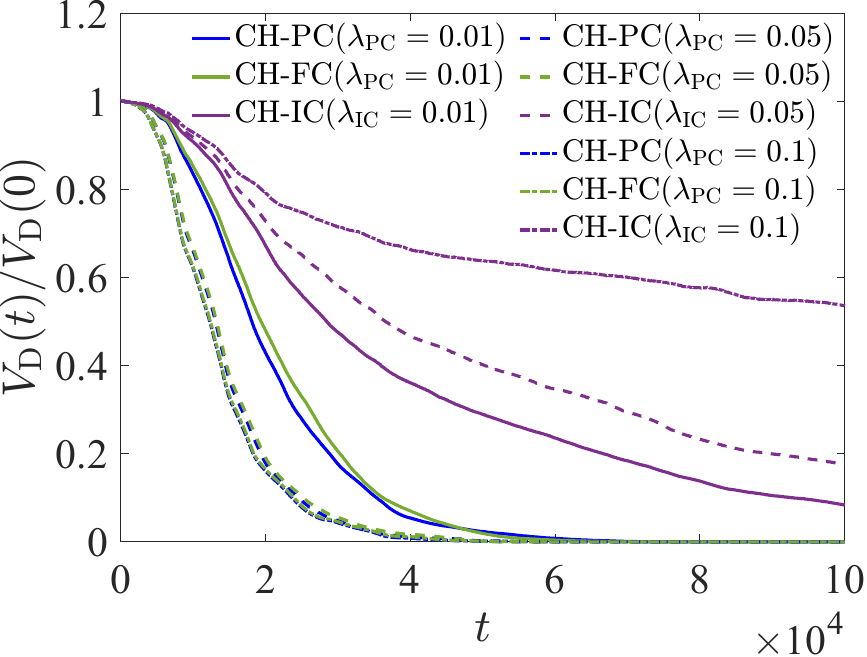}
    \caption{Comparison of droplet volume evolution under different tunable parameters for the CH-PC, CH-FC, and CH-IC models.  }
    \label{fig:WE20_DropletVolume_different_lambda}
\end{figure}

For the CH-IC model, we varied $\lambda_{\rm IC}$ over $[0.01, 0.1]$ and found that larger values improve droplet volume preservation, although further increases lead to numerical instability, consistent with the original findings of Zhang et al.~\cite{zhang2019interface}. Figure~\ref{fig:WE20_DropletVolume_different_lambda} presents the temporal evolution of droplet volume for the three modified CH models under three representative values of the tuning parameters: 0.01, 0.05, and 0.1. These results confirm that the relatively poor performance observed in Figure~\ref{fig:WE20_DropletVolume_6Models} is not attributable to parameter selection.

\begin{figure}
    \centering
    \includegraphics[width=0.5\textwidth]{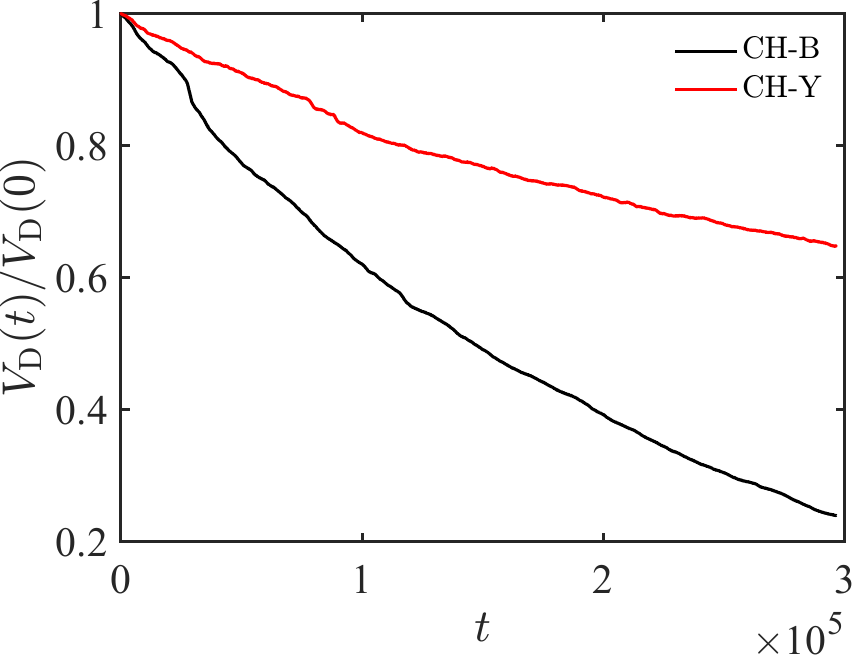}
    \caption{The evolution of droplet volume for the two modified CH models with singular mobility, CH-B and CH-Y. }
    \label{fig:WE1_2_DropletVolume_BandY}
\end{figure}

\begin{figure}

\begin{subfigure}[b]{0.45\textwidth}
        \centering
        \includegraphics[width=\textwidth]{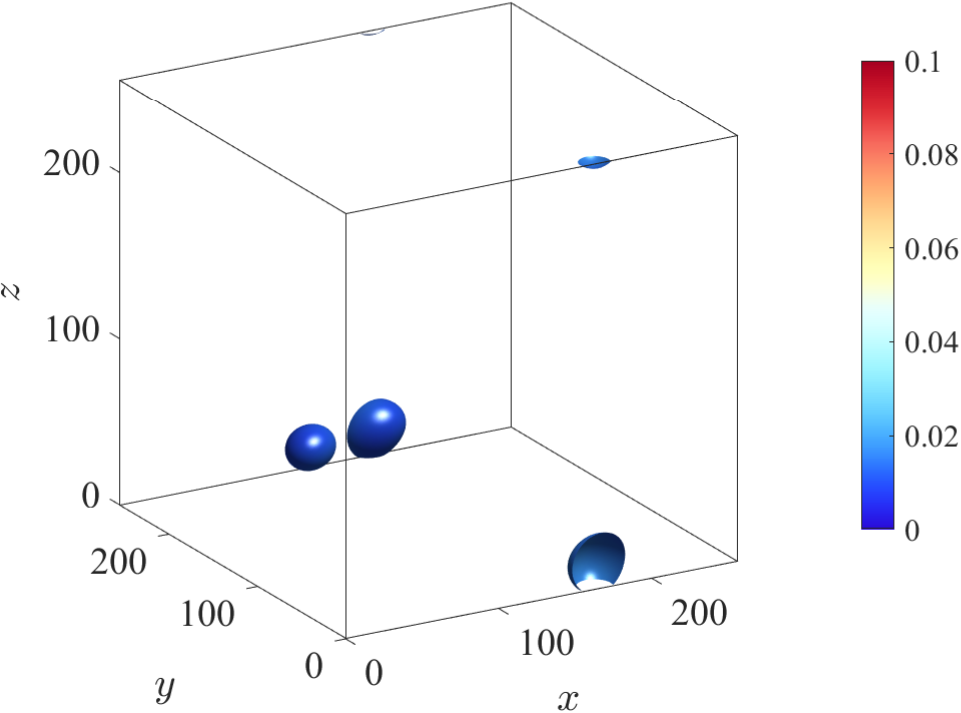}
        \caption{CH-B}
        \label{fig:visualizations of Bao}
\end{subfigure}
\hfill
    \begin{subfigure}[b]{0.45\textwidth}
        \centering
        \includegraphics[width=\textwidth]{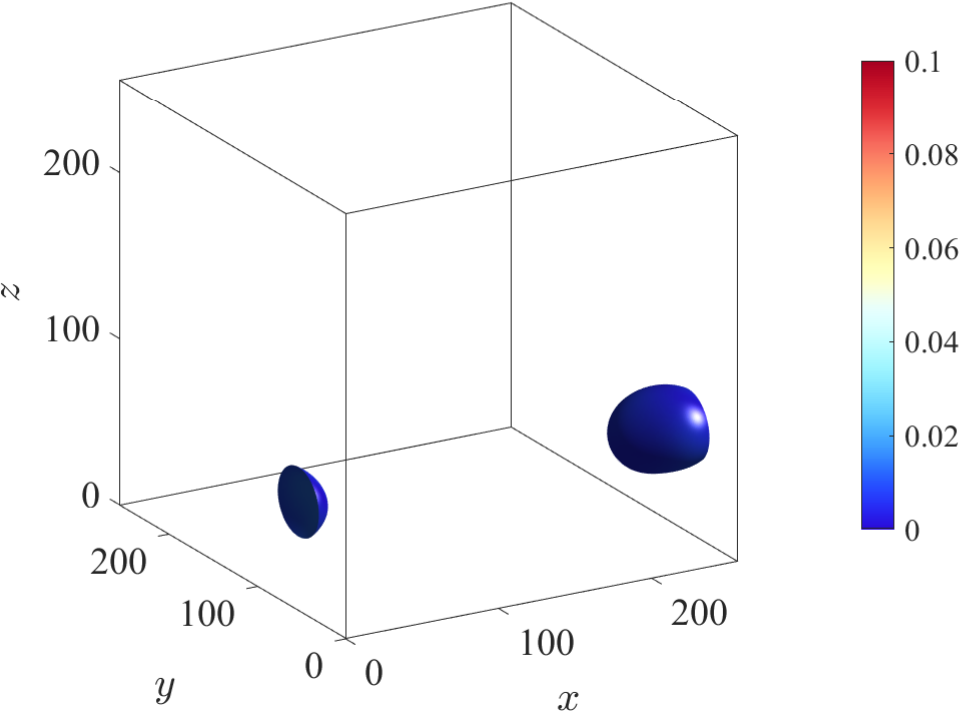}
        \caption{CH-Y}
        \label{fig:visualizations of Yang}
    \end{subfigure}

    \caption{Visualizations of the two-phase interface at time step $250,000$ at ${\rm We}_{\rm d} = 1.2$.}
\label{fig:visualizations of Bao and Yang}
\end{figure}

Next, we consider the CH-B and CH-Y models, which employ singular mobility. These models are discussed separately because their numerical stability is weaker than that of the previously examined five models. At the Weber number ${\rm We}_{\rm d} = 20$ used in the earlier tests, both models exhibited numerical instabilities. To enable long-term simulations, we therefore reduced the Weber number to ${\rm We}_{\rm d} = 1.2$, while keeping all other parameters identical.
The tuning parameter in the CH-Y model is set to $\gamma = 10^4$. Our observations indicate that CH-B demonstrates superior numerical stability, whereas CH-Y is more effective at preserving droplet volume. Figure~\ref{fig:WE1_2_DropletVolume_BandY} shows the temporal evolution of droplet volume for the two models. Despite these differences, both models still exhibit a gradual decrease of droplet volume over time. Figure~\ref{fig:visualizations of Bao and Yang} presents snapshots of the droplet interfaces at time step 250,000. Even at this low Weber number, where droplet breakage is limited, a significant shrinkage of droplet volume is apparent.

Finally, we examine the global mass-correction model of Wang et al.~\cite{wang2015mass}. To avoid a singularity in the correction term, which contains $\Delta \rho = \rho_{\rm D} - \rho_{\rm C}$ in the denominator, we set $\rho_{\rm D} = 2.0$ and $\rho_{\rm C} = 1.0$, while keeping all other parameters unchanged.

\begin{figure}

\begin{subfigure}[b]{0.45\textwidth}
    \centering
    \includegraphics[width=\textwidth]{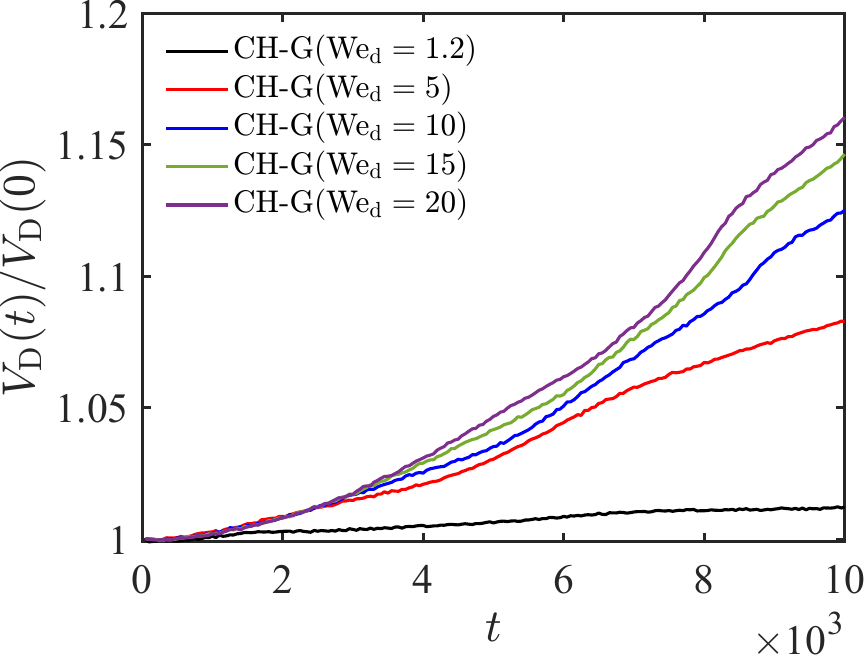} 
    \caption{Droplet volume}
    \label{fig:droplet_volume_evolution}
\end{subfigure}
\hfill
\begin{subfigure}[b]{0.45\textwidth}
    \centering
    \includegraphics[width=\textwidth]{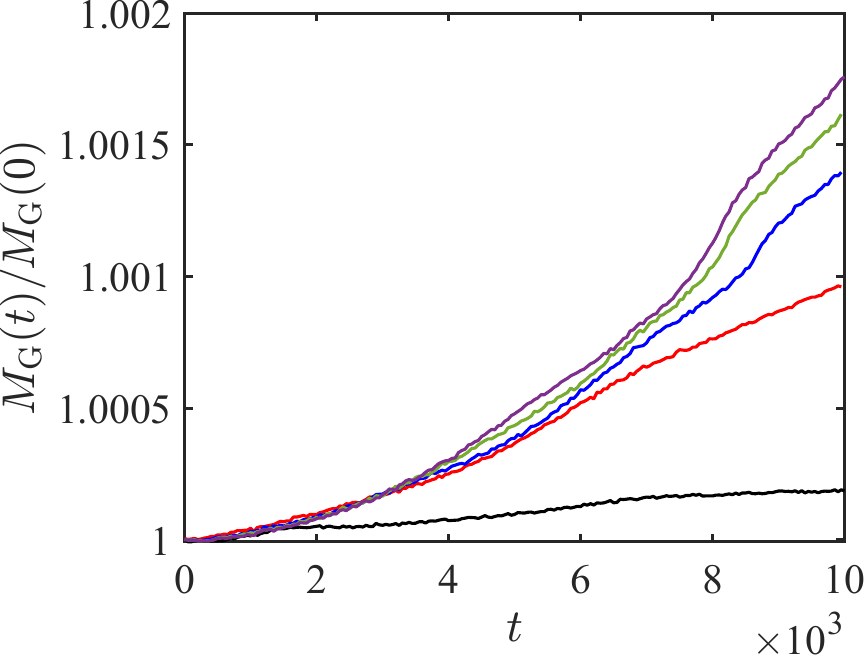} 
    \caption{Total mass in the system}
    \label{fig:total_mass_evolution}
\end{subfigure}

\caption{Evolution of droplet volume and total mass in the system with the global mass-correction model under different Weber numbers.}
\label{fig:global_mass_correction_evolution}
\end{figure}

Figure~\ref{fig:global_mass_correction_evolution} shows the evolution of droplet volume and total phase-field mass, $\int_\Omega \phi dV$, for this CH-G model. The results indicate that the model fails to maintain a constant droplet volume and also violates overall mass conservation. This limitation arises primarily from the inaccuracy of droplet volume evaluation using the $\phi \ge \phi_0$ criterion. As discussed in detail in Appendix~\ref{sec:appendixB}, for phase-field methods with diffused interfaces, strictly enforcing instantaneous droplet volume under conditions of significant interface deformation is not meaningful, because a volume bias is inherently introduced by the diffuse interface. The bias increases with the Weber number, as the total interfacial area grows when the initially large droplet is stretched and fragmented by turbulence, which is consistent with the trend observed in Figure~\ref{fig:global_mass_correction_evolution}. Furthermore, since this model only compensates for dispersed-phase mass lost to the carrier phase without removing the excess dissolved mass, the total system mass gradually increases over time.

\begin{figure}[htbp]
    \centering
    \begin{subfigure}[b]{0.32\textwidth}
        \centering
        \includegraphics[width=\textwidth]{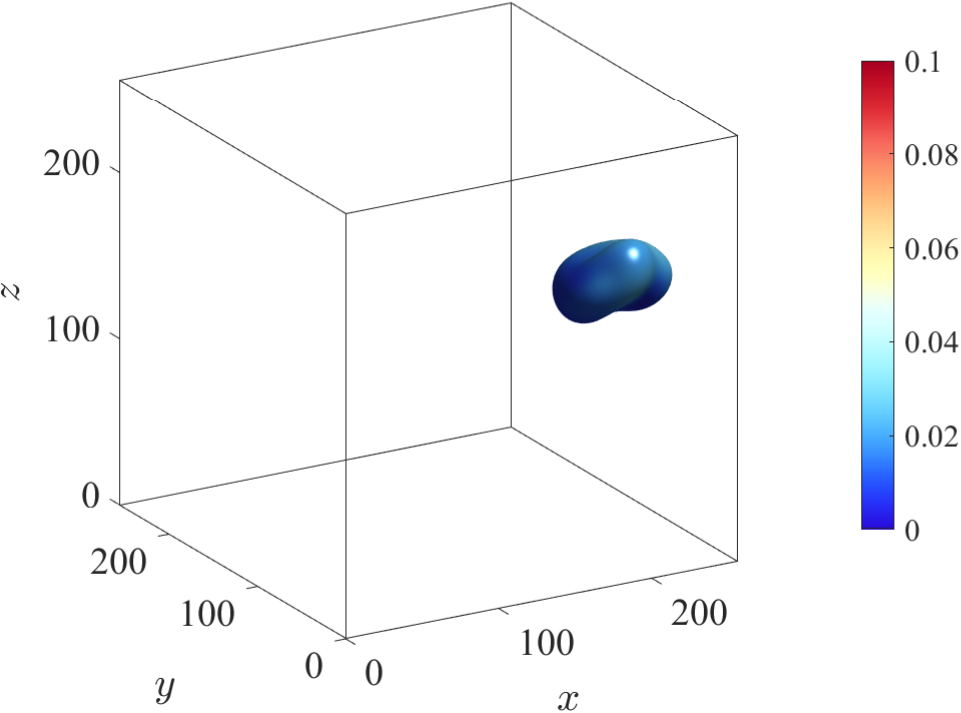}
        \caption{${\rm We}_{\rm d}=1.2$}
        \label{fig:GMC_We1_2}
    \end{subfigure}
    \hfill
    \begin{subfigure}[b]{0.32\textwidth}
        \centering
        \includegraphics[width=\textwidth]{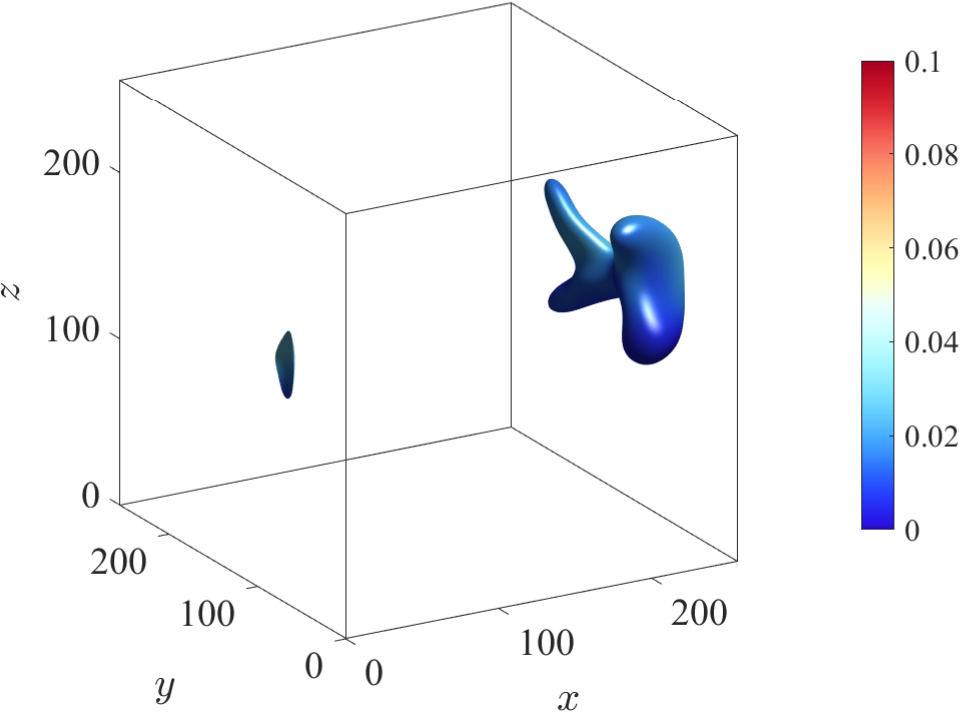}
        \caption{${\rm We}_{\rm d}=5$}
        \label{fig:GMC_We5}
    \end{subfigure}
    \hfill
    \begin{subfigure}[b]{0.32\textwidth}
        \centering
        \includegraphics[width=\textwidth]{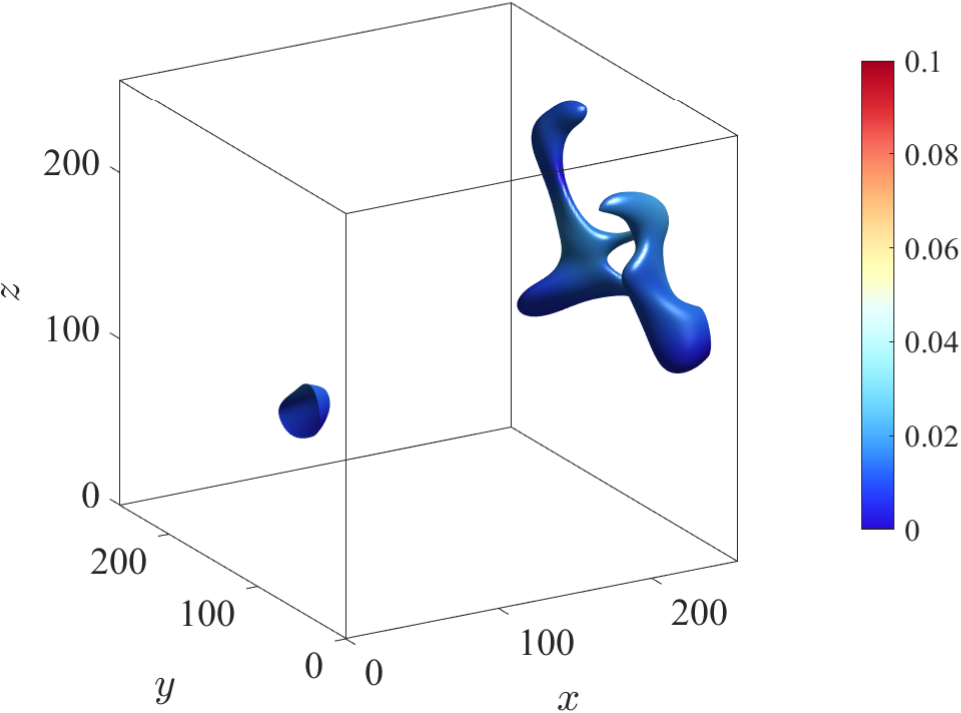}
        \caption{${\rm We}_{\rm d}=10$}
        \label{fig:GMC_We10}
    \end{subfigure}

    \vspace{1em} 

    \begin{subfigure}[b]{0.32\textwidth}
        \centering
        \includegraphics[width=\textwidth]{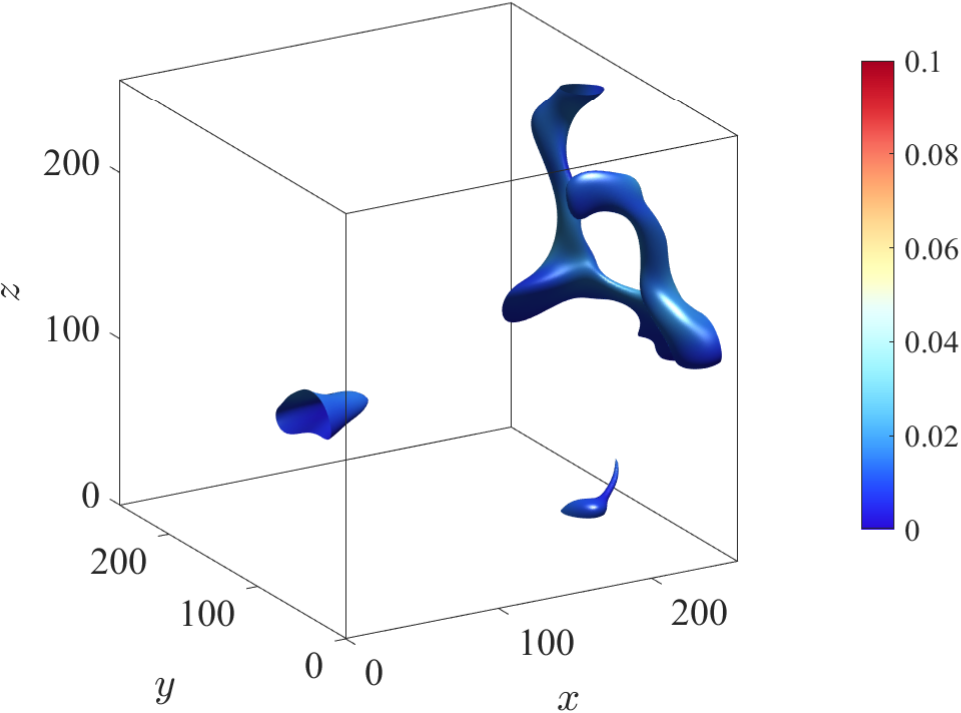}
        \caption{${\rm We}_{\rm d}=15$}
        \label{fig:GMC_We15}
    \end{subfigure}
    \hspace{0.04\textwidth}
    \begin{subfigure}[b]{0.32\textwidth}
        \centering
        \includegraphics[width=\textwidth]{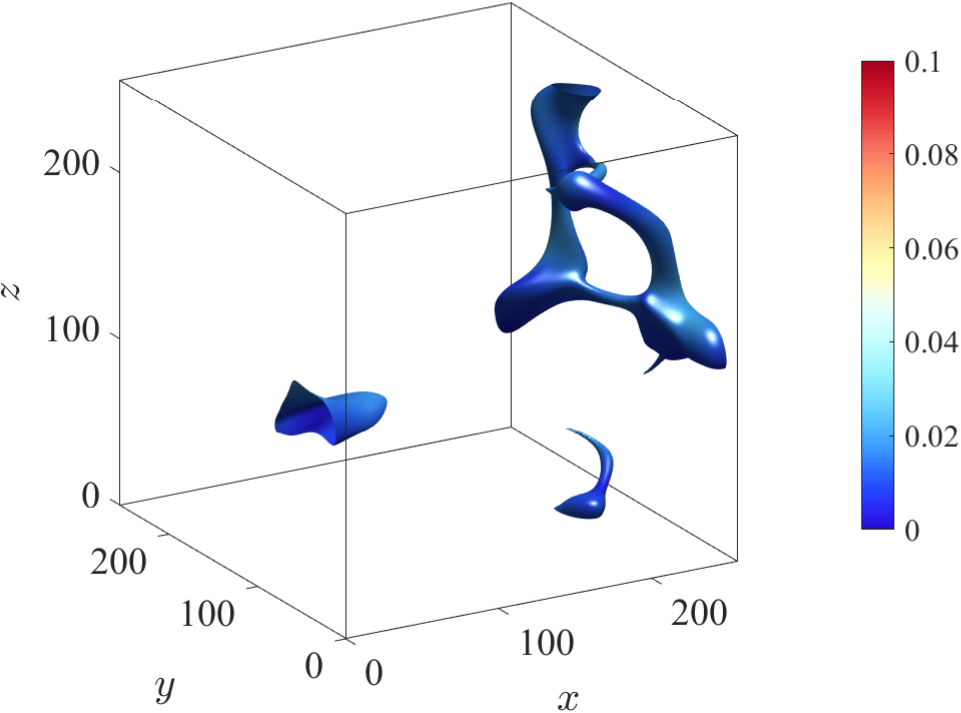}
        \caption{${\rm We}_{\rm d}=20$}
        \label{fig:GMC_We20}
    \end{subfigure}

    \caption{Visualizations of the two-phase interface for the global mass-correction model at different Weber numbers.}
    \label{fig:GMC_5We}
\end{figure}

Figure~\ref{fig:GMC_5We} presents snapshots of the interface for the CH-G model at different Weber numbers. Small droplets are largely absent, while the initial droplet progressively enlarges, indicating the occurrence of artificial coarsening.

\section{A modified CAC equation for droplet volume conservation}\label{sec:modifiedCAC}

Unlike laminar flows, droplet-laden turbulence is characterized by random velocity fluctuations and a broad spectrum of droplet sizes spanning multiple length scales. This intrinsic multiscale nature makes it extremely difficult to regulate the dissolution of individual droplets. Continuous breakup and coalescence processes incessantly generate small droplets whose sizes may become comparable to the interface thickness; such droplets are particularly susceptible to rapid dissolution into the carrier phase in phase-field simulations. As a result, some degree of droplet dissolution is essentially unavoidable in phase-field simulations of turbulent flows.

Motivated by this observation, our strategy is to introduce a counter-diffusion term acting along the interface normal direction. This term acts as a precipitation-like mechanism that selectively counteracts the excessive dissolution of small droplets caused by turbulent deformation and numerical diffusion under prescribed conditions.  By carefully controlling its strength, droplet volume can be conserved in a statistical sense, even without enforcing instantaneous conservation for each individual droplet.

The counter-diffusion term is designed to satisfy the following requirements:
\begin{itemize}
    \item First, it should affect only droplet volume conservation without altering the total mass or momentum of the system. Accordingly, the correction must be incorporated into the phase-field equation exclusively in divergence form, ensuring that its integral over the entire domain vanishes.
    \item Second, the correction must activate only for small droplets with large curvature. For flat interfaces or sufficiently large droplets, where dissolution is negligible, it should remain inactive.
    \item Third, the correction must avoid inducing artificial coarsening, i.e., any unphysical mass transfer from smaller droplets to larger ones. The objective is to preserve small droplets within the turbulent field, not to cause their absorption by larger ones.
\end{itemize}
In addition, the correction should be computationally efficient and should not introduce significant additional computational or communication overhead in parallel implementations.

Based on these considerations, we propose the following modified CAC equation:
\begin{equation}
    \frac{\partial \phi}{\partial t} + \boldsymbol{\nabla} \cdot (\phi \bm{u}) 
=  \boldsymbol{\nabla} \cdot \left\{M_{\rm AC}\left[ \boldsymbol{\nabla} \phi + \frac{4}{W} \frac{(\phi - \phi_{\rm D})(\phi - \phi_{\rm C})}{\phi_{\rm D} - \phi_{\rm C}}{\bm n} \right]\right\} - a\boldsymbol{\nabla}\cdot{\bm n},
\end{equation}
where the last term represents the counter-diffusive correction. This term is proportional to the interface curvature, with its strength controlled by the coefficient $a$ ($a>0$). When $a$ is chosen as a global constant, the correction remains in divergence form, ensuring that its integral over the entire computational domain vanishes under natural boundary conditions and leaving the total system mass unchanged. Furthermore, because the correction magnitude scales with curvature, its influence is strongest 
on small droplets and negligible for flat interfaces.

According to Hinze’s law~\cite{hinze1955fundamentals}, as the Weber number increases, the maximum droplet size that can stably exist in a turbulent flow without breakup decreases. Consequently, the strength of the counter-diffusion term should increase with increasing Weber number. We therefore propose the following empirical relation for $a$:
\begin{equation}
    a = \alpha M_{\rm AC} \frac{\phi_{\rm D} - \phi_{\rm C}}{d_{\rm crit}},
\end{equation}
where 
\begin{equation}
   d_{\rm crit} = \frac{\sigma {\rm We}}{0.725\rho_{\rm C}u'^2 }
\end{equation}
is the critical droplet diameter predicted by Hinze’s law, above which droplets are expected to undergo breakup in a turbulent flow. Here, $u'$ denotes the root-mean-square velocity fluctuation of the turbulent field, and $\alpha$ is a dimensionless parameter to be determined empirically. 

By linking $a$ to $u'$ through Hinze’s law, the counter-diffusion term is naturally adapted for turbulence-dominated regimes. In laminar flows, where $u' = 0$, this relation yields $a = 0$, causing the correction term to vanish. This is consistent with the observation that the original CAC formulation is generally adequate for laminar flows and for turbulent cases at sufficiently low Weber numbers, where no additional counter-diffusive correction is required~\cite{li2025comparativestudycriticalassessment}. If needed, however, the parameter $a$ can be directly prescribed in laminar simulations to preserve droplet volume. In general, increasing $a$ strengthens the barrier against the dissolution of small droplets, while decreasing $a$ weakens this effect.

In practice, the proposed correction can be absorbed into the existing CAC correction term, yielding the following modified CAC equation:
\begin{equation}
  \frac{\partial \phi}{\partial t} + \boldsymbol{\nabla} \cdot (\phi \bm{u}) 
=  \boldsymbol{\nabla} \cdot \left\{M_{\rm AC}\left[ \boldsymbol{\nabla} \phi + \left(\frac{4}{W} \frac{(\phi - \phi_{\rm D})(\phi - \phi_{\rm C})}{\phi_{\rm D} - \phi_{\rm C}}-\alpha\frac{\phi_{\rm D}-\phi_{\rm C}}{d_{crit}} \right){\bm n} \right]\right\}.
\end{equation}
As a result, the modified formulation does not introduce any additional computational cost.

The detailed LB implementation of the modified CAC equation is provided in Appendix~\ref{sec:appendixA}. As demonstrated in the previous section and further confirmed in the next section, this modified CAC equation enables statistical conservation of droplet volume over sufficiently long simulation times required for the collection of turbulent statistics.

\section{Numerical validations}\label{sec:validation}

In this section, we demonstrate the effectiveness of the proposed modified CAC equation in preserving droplet volume in DNS of droplet-laden HIT.  This test corresponds to the same configuration in which all previously introduced phase-field methods failed, as discussed in Sec.~\ref{sec:previous}. Owing to the high computational cost of such simulations, the proposed method is evaluated for a limited number of representative cases, with the corresponding parameters summarized in Table~\ref{tab:cases}.

\begin{figure}
\begin{subfigure}[b]{0.24\textwidth}
        \centering
        \includegraphics[width=\textwidth]{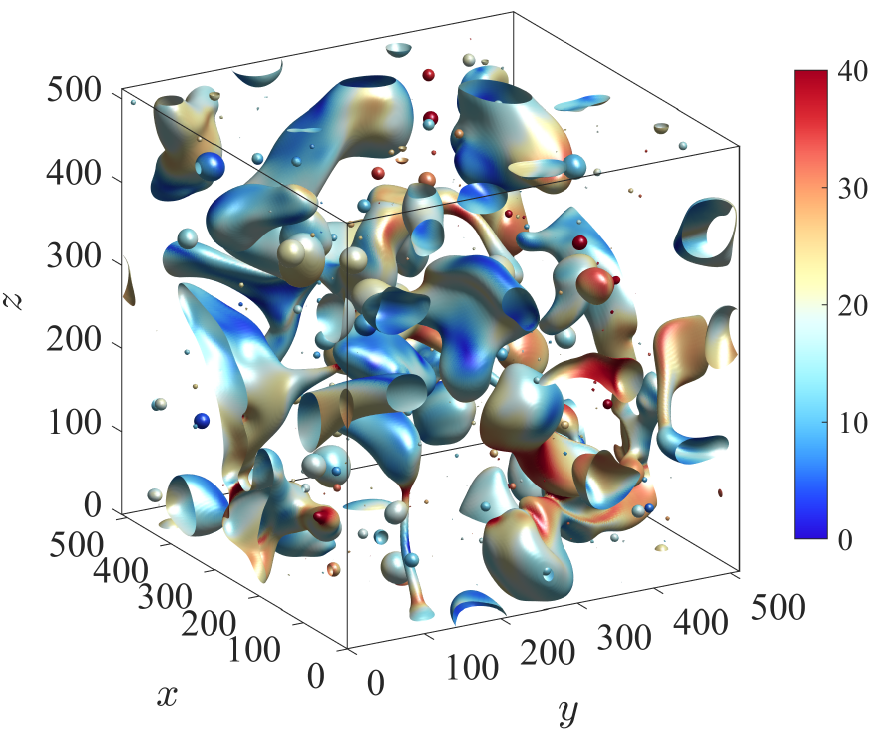}
        \caption{Case 1}
        \label{fig:isosurface1}
    \end{subfigure}
    \hfill
    \begin{subfigure}[b]{0.24\textwidth}
        \centering
        \includegraphics[width=\textwidth]{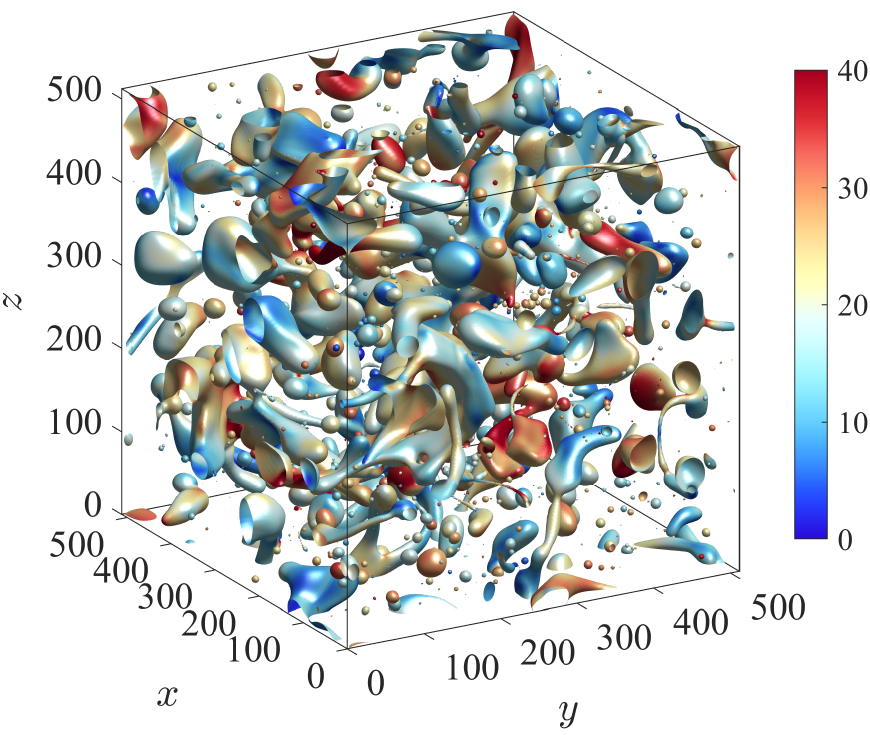}
        \caption{Case 2}
        \label{fig:isosurface2}
    \end{subfigure}
   \hfill
    \begin{subfigure}[b]{0.24\textwidth}
        \centering
        \includegraphics[width=\textwidth]{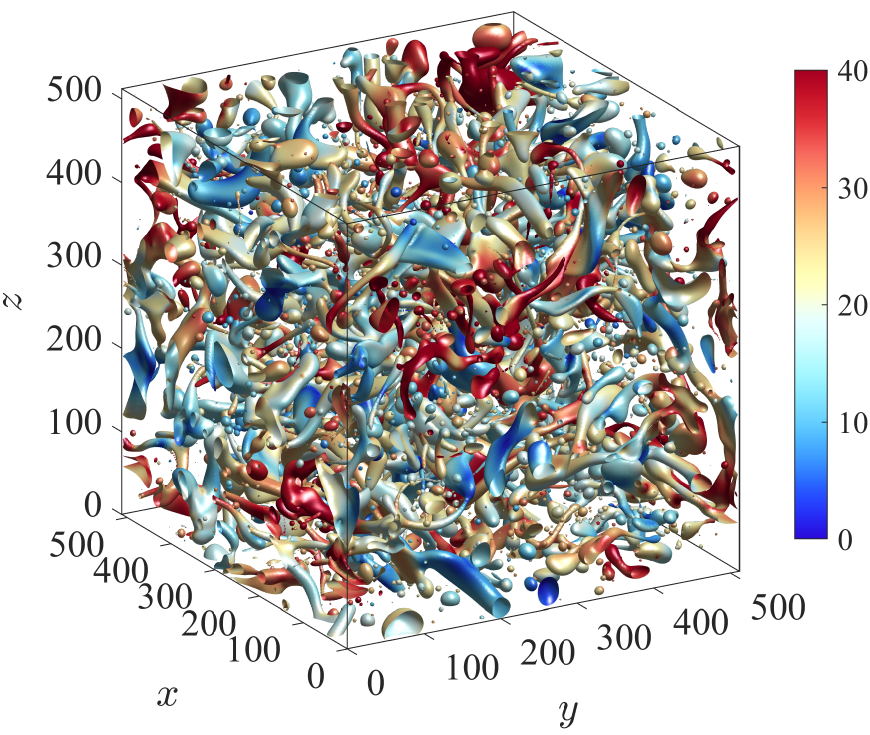}
        \caption{Case 3}
        \label{fig:isosurface3}
    \end{subfigure}
    \hfill
    \begin{subfigure}[b]{0.24\textwidth}
        \centering
        \includegraphics[width=\textwidth]{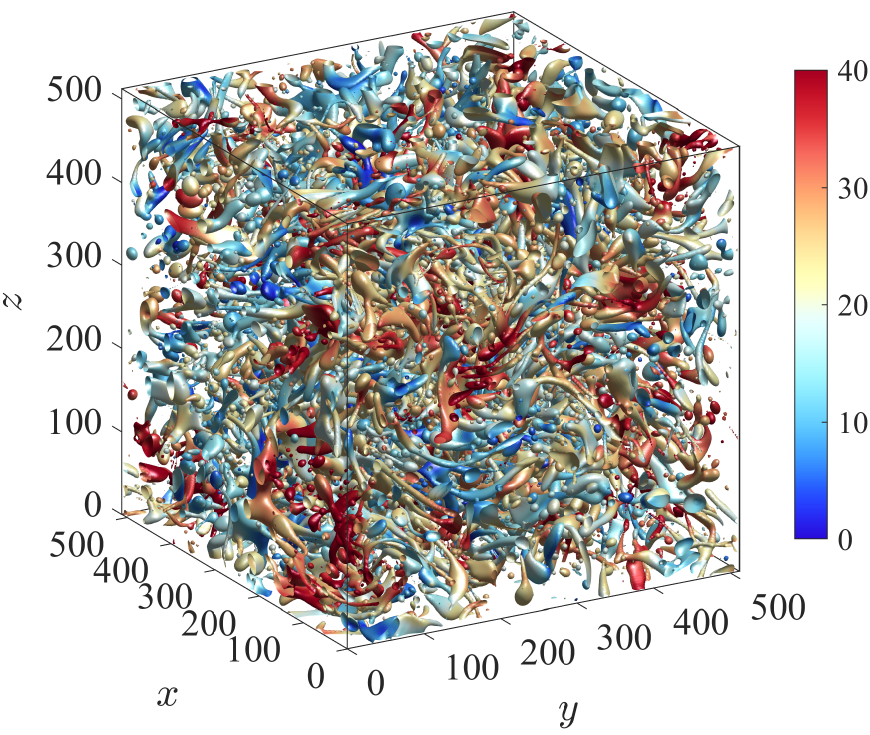}
        \caption{Case 4}
        \label{fig:isosurface4}
    \end{subfigure}
    \vspace{0.5em}

    \qquad\qquad\qquad\begin{subfigure}[b]{0.24\textwidth}
        \centering
        \includegraphics[width=\textwidth]{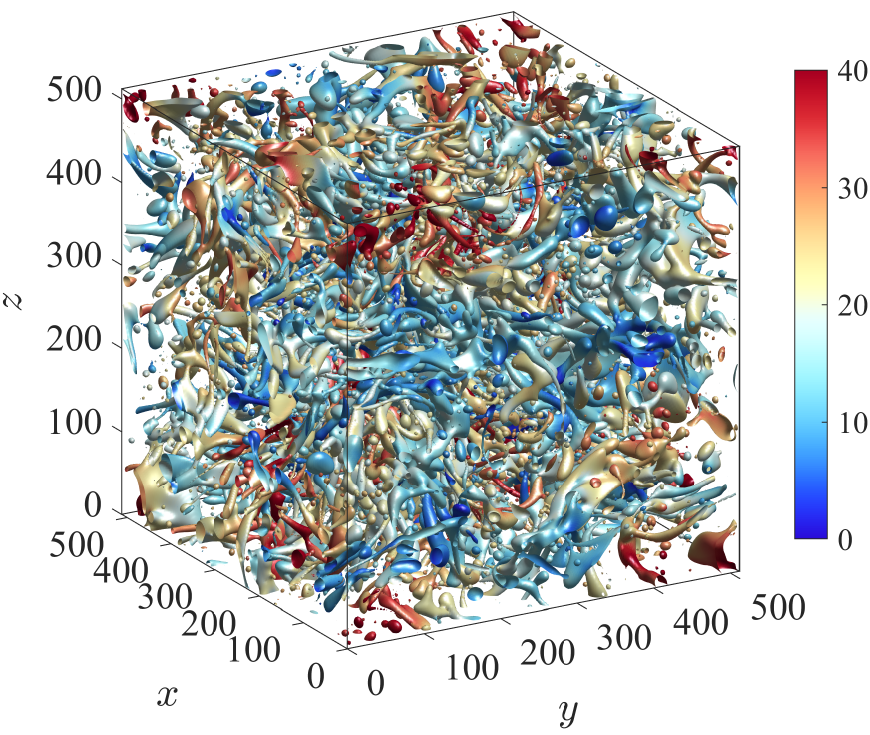}
        \caption{Case 5}
        \label{fig:isosurface5}
    \end{subfigure}
    \hfill
    \begin{subfigure}[b]{0.24\textwidth}
        \centering
        \includegraphics[width=\textwidth]{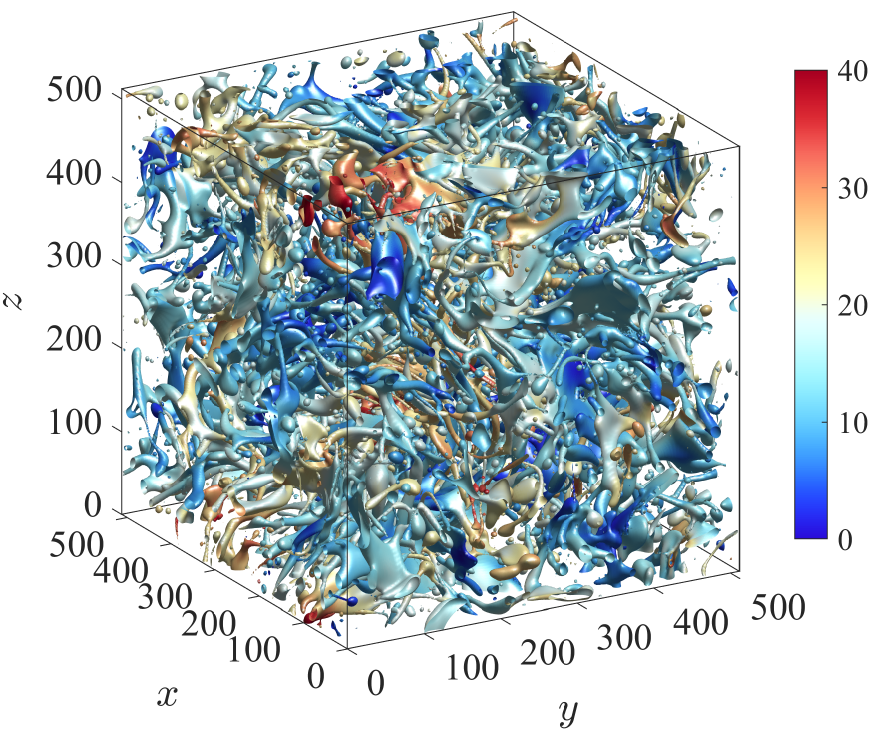}
        \caption{Case 6}
        \label{fig:isosurface6}
    \end{subfigure}
    \hfill
    \begin{subfigure}[b]{0.24\textwidth}
        \centering
        \includegraphics[width=\textwidth]{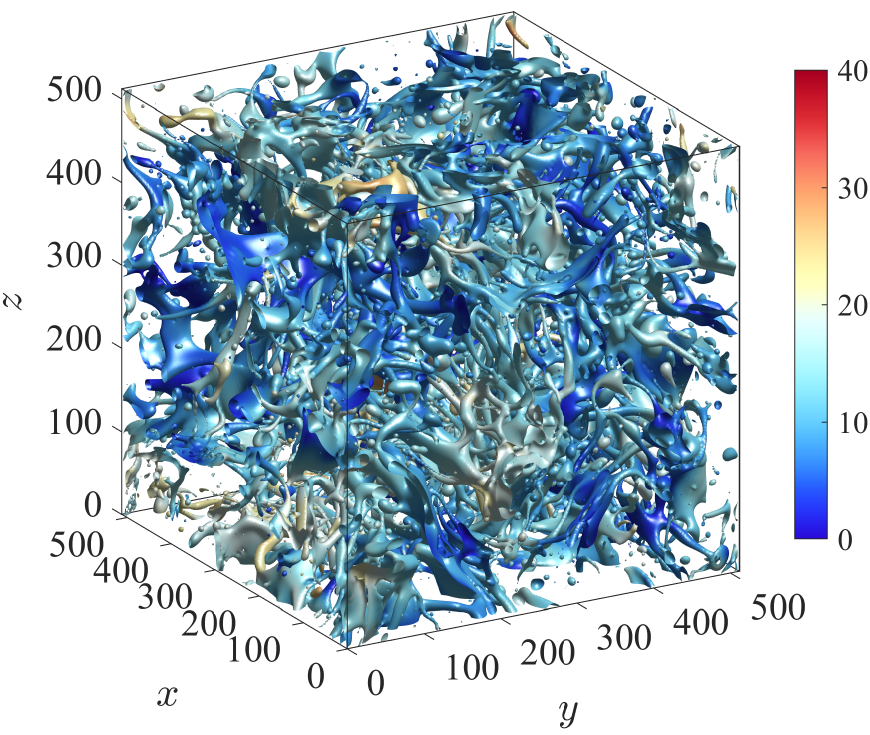}
        \caption{Case 7}
        \label{fig:isosurface7}
    \end{subfigure}\qquad\qquad\qquad

\caption{Instantaneous droplet distribution in the droplet-laden turbulence DNS. The two-phase interface is identified with $\phi = 0.5$ and the color represents the velocity magnitude (in spectral unit). }
\label{fig:isosurfaces}
\end{figure}

A total of seven droplet-laden turbulent cases are simulated. Cases 1–4 employ unity density and viscosity ratios, while the Weber number—defined based on the Taylor microscale as ${\rm We} = \rho_{\rm C} u'^2 \lambda / \sigma$—is varied from 1 to 10. Using the droplet-based definition ${\rm We}_{\rm d}$, these values correspond to ${\rm We}_{\rm d} \approx 4.2$–42, respectively.
Cases 4–7 maintain the same ${\rm We} = 10$ but vary the density ratio from 1 to 10. For the first four cases, a constant value of $\alpha = 0.037$ is adopted, under which the proposed method successfully preserves the droplet volume in a statistical sense. In contrast, the parameter $\alpha$ is adjusted for the last three cases for two main reasons. First, changes in viscosity can influence the numerical dissolution rate of droplets. Second, the introduction of droplets with different densities alters the system inertia, leading to variations in the turbulent velocity fluctuations $u'$ (see the different color scales in Figure~\ref{fig:isosurfaces}) and, consequently, in the effective Weber number. Under such conditions, the critical diameter $d_{\rm crit}$ estimated from single-phase flow properties may no longer be accurate, necessitating further tuning of $\alpha$.

All simulations are initialized following the same procedure described in Section~\ref{sec:previous}, and the statistical properties of the background turbulence are listed in Table~\ref{tab:statistics}. To improve the reliability of the physical results, all simulations are performed on a uniform mesh with a resolution of $512^3$. The interface thickness is fixed at $W = 3\delta x$, and the mobility is set to $M_{\rm AC} = 0.01$, with $\phi_{\rm D} = 1$ and $\phi_{\rm C} = 0$. The carrier-phase density is fixed at $\rho_{\rm C} = 1.0$ in lattice Boltzmann units. The kinematic viscosity of the carrier phase, $\nu_{\rm C}$, is set to $0.006$ (in $\delta x^2/\delta t$) for Cases 1–4 and increased to 0.01 for Cases 5–7 to enhance numerical stability. The corresponding properties of the dispersed phase, $\rho_{\rm D}$ and $\nu_{\rm D}$, are determined according to the prescribed density and viscosity ratios.

 \begin{table}[h]
    \centering
    \caption{The parameter setting in droplet-laden DNS. The items from the second to the last column are: the Weber number, the density ratio, the viscosity ratio, the droplet volume fraction, the value of the tuning parameter.}
    \label{tab:cases}
    \begin{tabular}{c c c c c c c c c}
        \hline
         &  We & $\rho_{\rm D}/\rho_{\rm C}$ & $\nu_{\rm D}/\nu_{\rm C}$ & $\varphi_{\rm D}$ & $\alpha$\\
        \hline
         Case 1 & 1.0 & 1.0 & 1.0 & 10\% & 0.037\\
         Case 2 & 2.0 & 1.0 & 1.0 & 10\% & 0.037\\
         Case 3 & 5.0 & 1.0 & 1.0 & 10\% & 0.037\\
         Case 4 & 10.0 & 1.0 & 1.0 & 10\% & 0.037\\
         Case 5 & 10.0 & 1.0 & 1.0 & 10\% & 0.041\\
         Case 6 & 10.0 & 1.0 & 1.0 & 10\% & 0.033\\
         Case 7 & 10.0 & 1.0 & 1.0 & 10\% & 0.029\\
        \hline
    \end{tabular}%
\end{table}

\begin{table}[h]
    \centering
    \caption{The statistics of the background turbulence in the spectral units with $\nu = 0.0945$ and $L = 2\pi$. The items in the table are: the root-mean-square velocity, the turbulent kinetic energy, the dissipation rate of turbulent kinetic energy, the Reynolds number defined with the Taylor-microscale $\lambda$, the Kolmogorov length, the skewness of longitudinal velocity gradient, the flatness of longitudinal velocity, the longitudinal integral length scale, the eddy turnover time, and Taylor-microscale. The uncertainties behind $\pm$ represents 95\% of confidence, i.e., $\sigma_U = 1.96\sigma_{A}\sqrt{2T_c(0.5)/T_{\rm ave}}$, where $\sigma_{A}$ is the standard deviation of the samples, $T_c(0.5)$ is the time difference corresponding to 0.5 of the autocorrelation, and $T_{\rm ave}$ is the averaging period~\cite{eswaran1988examination,wang2014study}.}
    \label{tab:statistics}
\begin{tabular}{c c c c c c c c c c c}
        \hline
         $u'$ & $E$ & $\epsilon$ & ${\rm Re}_\lambda$ & $\eta$\\
        \hline
         $13.93\pm0.10$&$291.52\pm4.26$&$1509.1\pm29.37$&$63.10\pm0.54$&$0.0274\pm0.0001$ \\
         \hline
         $S$&$F$&$L_{11}$&$T_{\rm e}$&$\lambda$\\
         \hline
$-0.503\pm0.002$&$4.669\pm0.019$&$1.081\pm0.008$&$0.130\pm0.001$&$0.428\pm0.002$\\
        \hline
    \end{tabular}%
\end{table}

Figure~\ref{fig:isosurfaces} presents snapshots of the droplet distribution after the simulations have reached statistically stationary states. As expected, increasing the Weber number enhances droplet deformation and rupture by turbulent fluctuations, leading to the breakup of large droplets into smaller ones. This process generates a substantial population of satellite droplets, whose number increases markedly with Weber number. As demonstrated in Section~\ref{sec:previous}, when the original CAC equation or various modified CH equations are employed, these small droplets dissolve rapidly into the carrier phase, resulting in a dramatic reduction of the total droplet volume over time. In contrast, when the proposed modified CAC equation is used, the total droplet volume remains statistically conserved.

\begin{figure}
\begin{subfigure}[b]{0.24\textwidth}
        \centering
        \includegraphics[width=\textwidth]{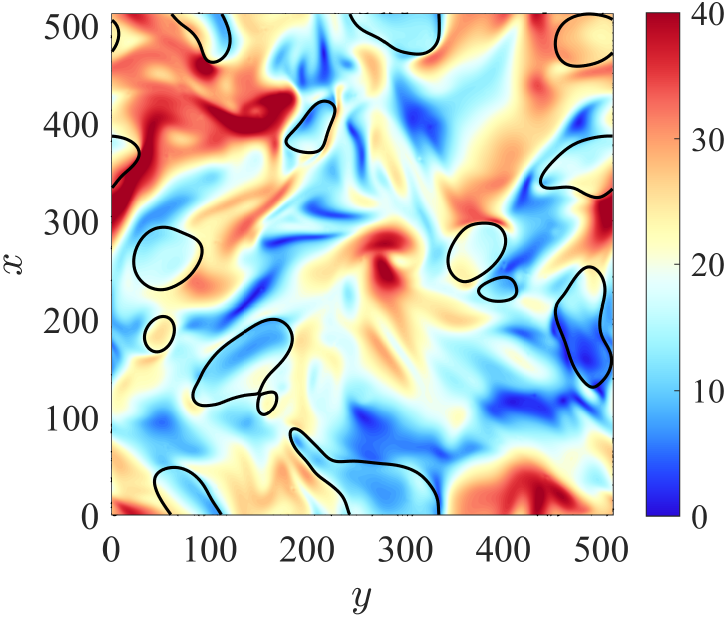}
        \caption{Case 1}
        \label{fig:contour1}
    \end{subfigure}
    \hfill
    \begin{subfigure}[b]{0.24\textwidth}
        \centering
        \includegraphics[width=\textwidth]{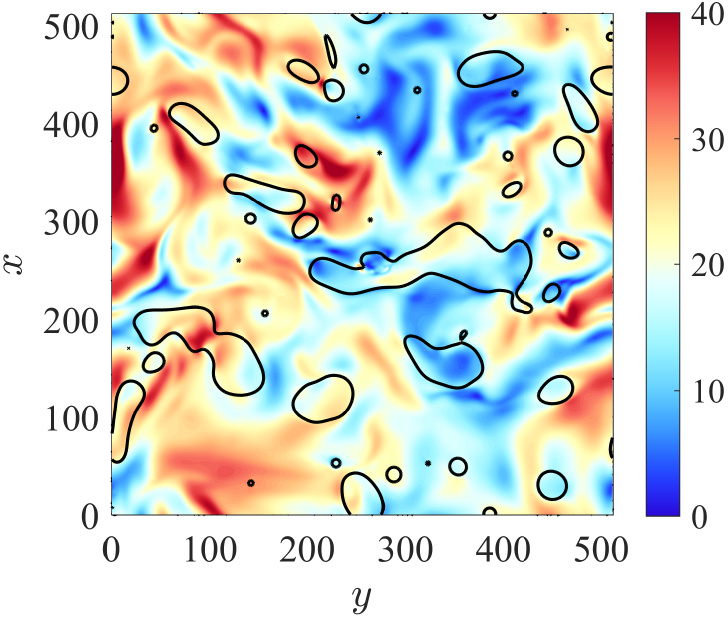}
        \caption{Case 2}
        \label{fig:contour2}
    \end{subfigure}
   \hfill
    \begin{subfigure}[b]{0.24\textwidth}
        \centering
        \includegraphics[width=\textwidth]{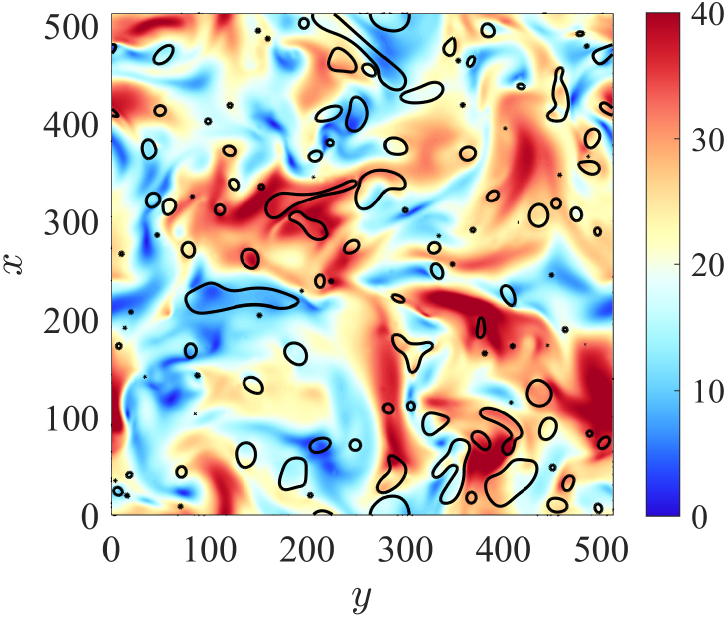}
        \caption{Case 3}
        \label{fig:contour3}
    \end{subfigure}
    \hfill
    \begin{subfigure}[b]{0.24\textwidth}
        \centering
        \includegraphics[width=\textwidth]{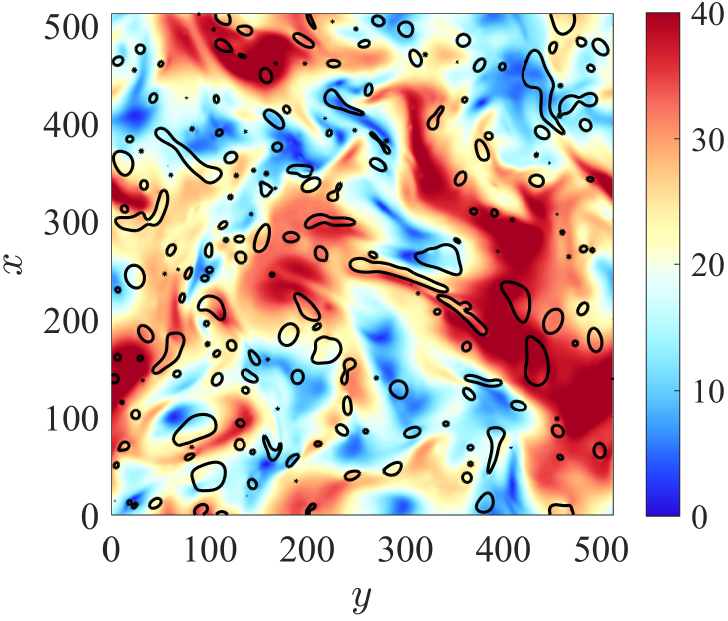}
        \caption{Case 4}
        \label{fig:Contour4}
    \end{subfigure}
    \vspace{0.5em}

    \qquad\qquad\qquad\begin{subfigure}[b]{0.24\textwidth}
        \centering
        \includegraphics[width=\textwidth]{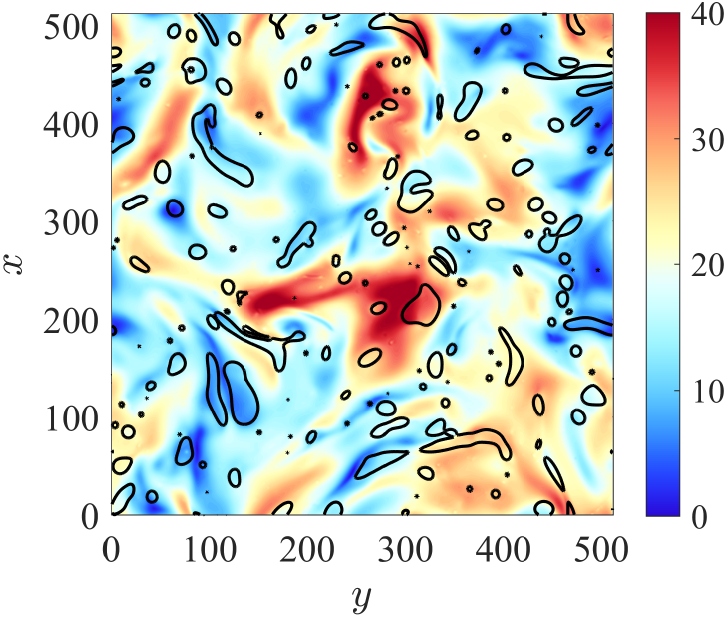}
        \caption{Case 5}
        \label{fig:Contour5}
    \end{subfigure}
    \hfill
    \begin{subfigure}[b]{0.24\textwidth}
        \centering
        \includegraphics[width=\textwidth]{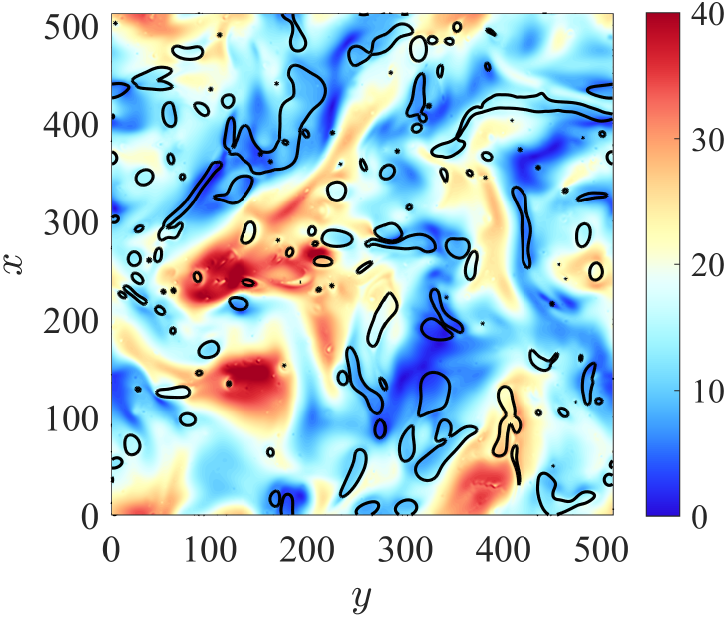}
        \caption{Case 6}
        \label{fig:Contour6}
    \end{subfigure}
    \hfill
    \begin{subfigure}[b]{0.24\textwidth}
        \centering
        \includegraphics[width=\textwidth]{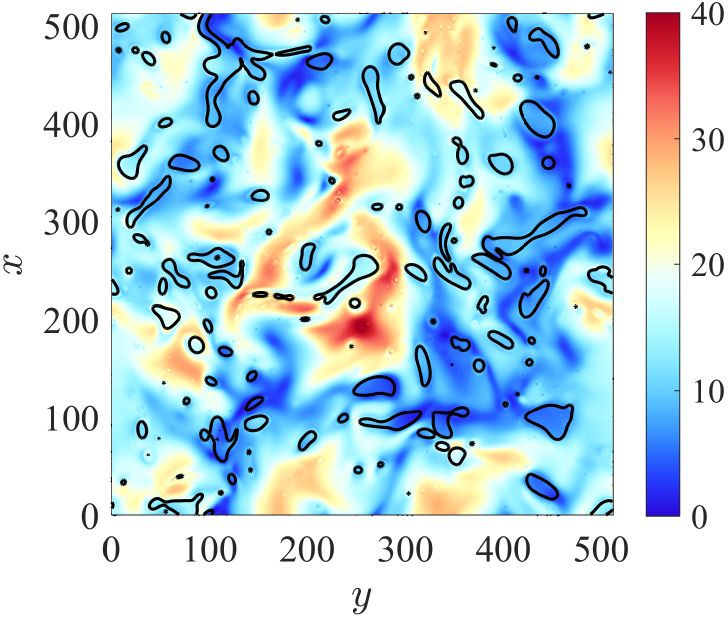}
        \caption{Case 7}
        \label{fig:Contour7}
    \end{subfigure}\qquad\qquad\qquad

\caption{Instantaneous droplet distribution on a cross section of the domain. The two-phase interface is identified with $\phi = 0.5$ and the color represents the velocity magnitude (in spectral unit). }
\label{fig:contours}
\end{figure}

To further illustrate the persistence of small droplets, Figure~\ref{fig:contours} shows a planar slice of the computational domain with droplets projected onto the selected plane. A significant number of small droplets with sizes comparable to the interface thickness are clearly visible. It is worth noting that several previous studies have reported apparently good droplet volume conservation using modified CH equations such as CH-PC and CH-FC. For example, Soligo et al.~\cite{soligo2019mass} reported droplet volume losses of only about 0.5\% for CH-FC and 2\% for CH-PC in simulations of droplet-laden turbulent channel flow. However, this apparent conservation relies on two specific conditions. First, the Weber numbers considered in those studies were relatively low, and as visualized therein, only a limited number of small droplets were generated—fundamentally different from the high-Weber-number regimes examined here. Second, the simulation times were relatively short. Since droplet volume loss is a continuous process, its cumulative effect remains small over short durations but can become substantial in long-time simulations, particularly under strong turbulent deformation.

\begin{figure}

\begin{subfigure}[b]{0.5\textwidth}
        \centering
        \includegraphics[width=\textwidth]{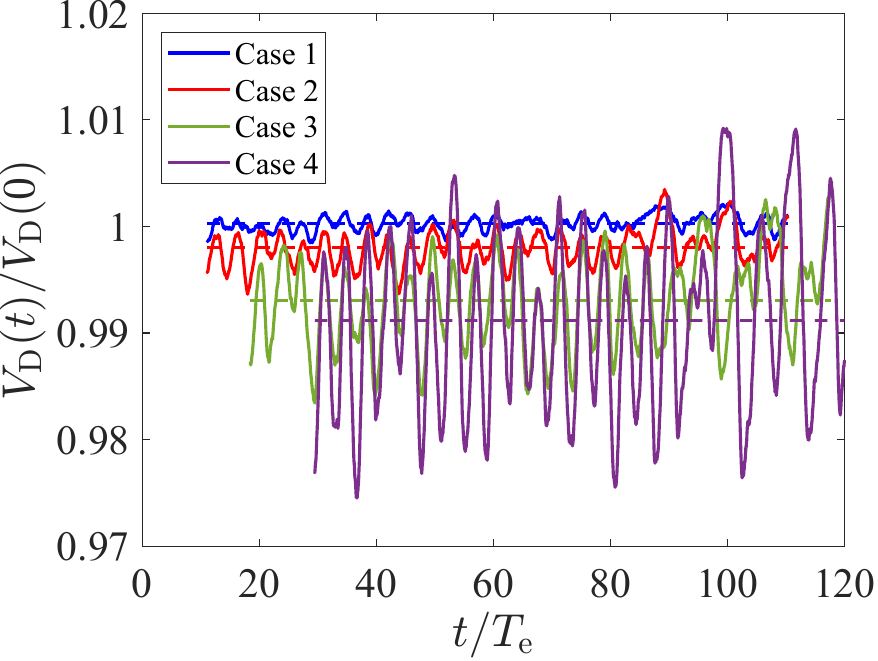}
        \caption{Case 1-4 with different We}
        \label{fig:volume1}
\end{subfigure}
\hfill
    \begin{subfigure}[b]{0.5\textwidth}
        \centering
        \includegraphics[width=\textwidth]{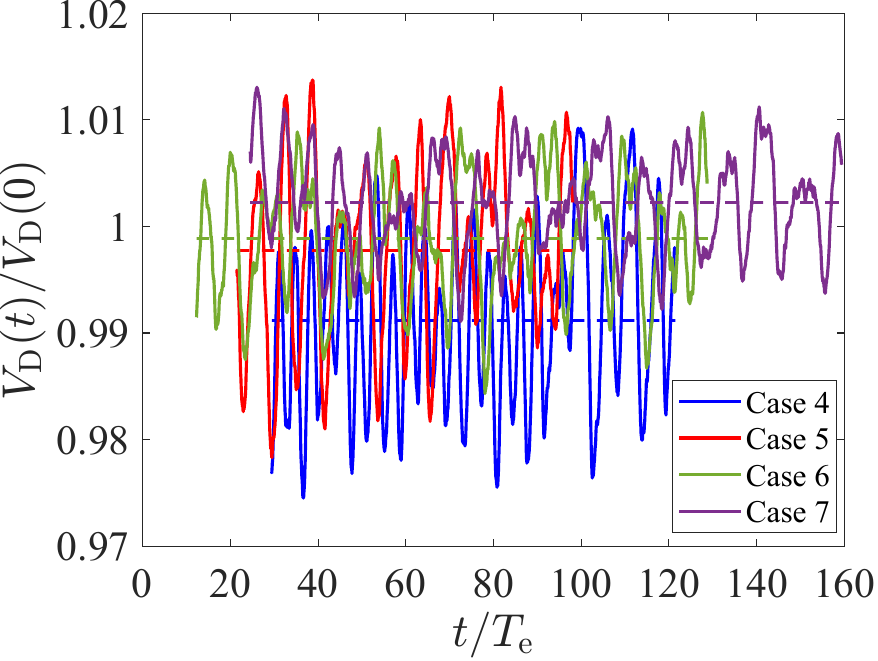}
        \caption{Case 4-7 with different density ratio}
        \label{fig:volume2}
    \end{subfigure}

    \caption{Time evolution of the droplet volume. $T_{\rm e}$ roughly corresponds to $1.36\times10^4$ time steps in Case 1-4, and $8.16\times10^3$ time steps in Case 5-7. The horizontal dash lines with the same color show the averaged volume fraction in each case.}
\label{fig:volume}
\end{figure}

The temporal evolution of the total droplet volume is shown in Figure~\ref{fig:volume}. Time is normalized by the eddy turnover time $T_{\rm e}$, corresponding to approximately $1.36\times10^{4}$ time steps in Cases~1–4 and $8.16\times10^{3}$ time steps in Cases~5–7, the latter being shorter due to the increased viscosity. All simulations were performed for more than $10^{6}$ time steps, over which a statistically stationary droplet volume is maintained.

For diffuse-interface methods such as the phase-field approach, droplet volume conservation is more appropriately assessed in a statistical sense when simulating droplet-laden turbulence. Accurate determination of the instantaneous droplet volume becomes extremely challenging in the presence of highly deformed interfaces, frequent breakup, and coalescence. In practice, the droplet volume is often estimated as the region where $\phi > 0.5$ (assuming $\phi_{\rm D}=1$ and $\phi_{\rm C}=0$). This criterion is inherently sensitive to interface morphology and therefore cannot yield strict volume conservation under strong deformation. As shown in Figure~\ref{fig:volume}, increasing the Weber number enhances droplet deformability, leading to progressively larger fluctuations in the measured volume (see analysis in Appendix~\ref{sec:appendixB}).

In this context, enforcing strict global volume conservation—such as in global correction approaches like that of Wang et al.~\cite{wang2015mass}—is neither practical nor physically meaningful, because the instantaneous droplet volume itself cannot be evaluated accurately. We emphasize that the proposed counter-diffusion correction does not eliminate droplet dissolution entirely. Instead, it allows dissolved droplets to precipitate under appropriate conditions, thereby compensating excessive volume loss and ensuring statistical conservation.

It is worth noting that the original CAC equation is already capable of preserving droplet volume in the absence of flow, even for arbitrarily small droplets, because the diffusive flux along the interface normal direction vanishes in a pure diffusion problem. Its failure in droplet-laden turbulence arises from the difficulty of maintaining this cancellation in the presence of strong, random velocity fluctuations, particularly in the process of droplet breakage. The additional counter-diffusion term introduced here is therefore essential for restoring statistical volume conservation under turbulent conditions.

\begin{figure}

\begin{subfigure}[b]{0.5\textwidth}
        \centering
        \includegraphics[width=\textwidth]{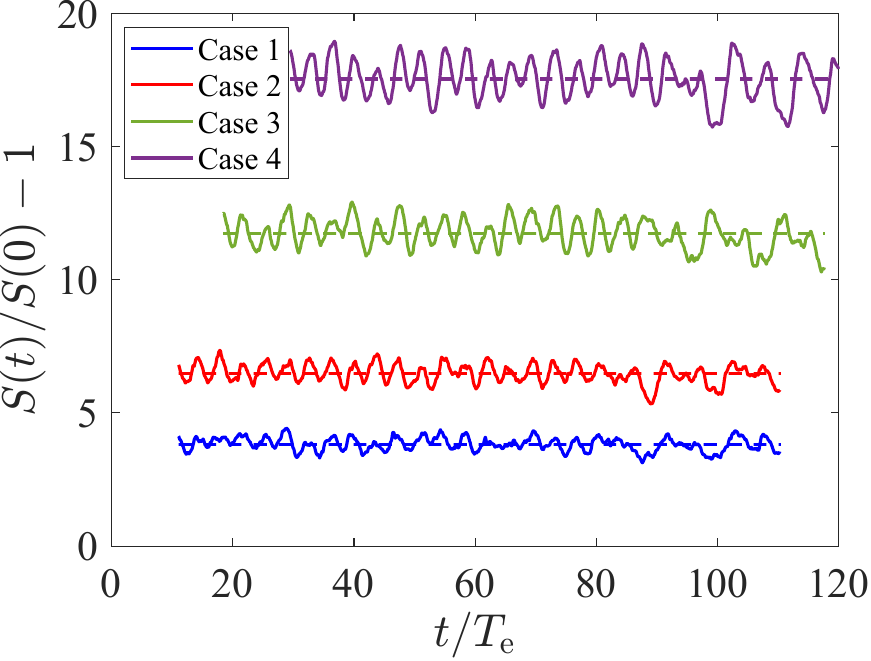}
        \caption{Case 1-4 with different We}
        \label{fig:surface1}
\end{subfigure}
\hfill
    \begin{subfigure}[b]{0.5\textwidth}
        \centering
        \includegraphics[width=\textwidth]{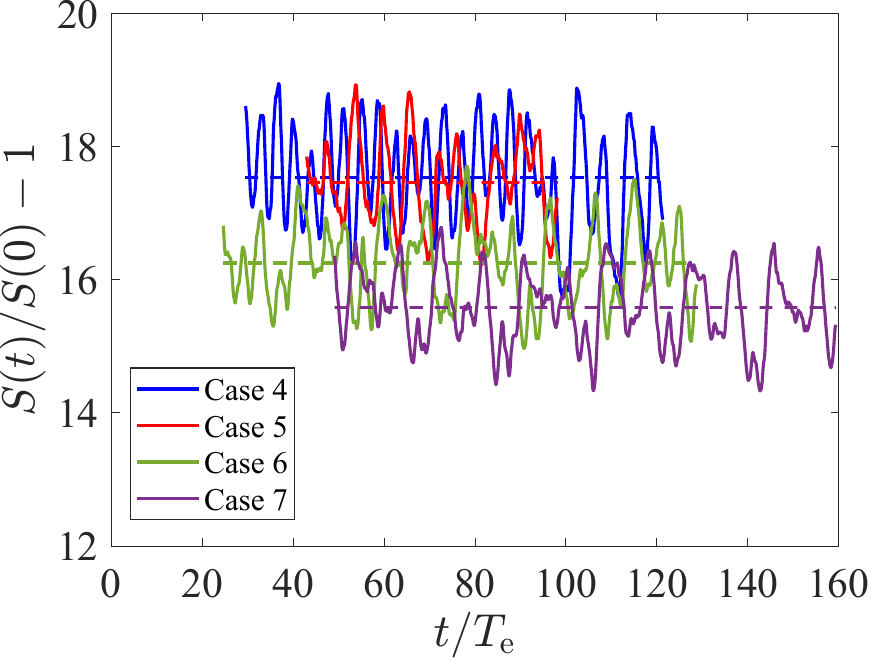}
        \caption{Case 4-7 with different density ratio}
        \label{fig:surface2}
    \end{subfigure}

    \caption{Time evolution of the interfacial area. The characteristic time $T_{\rm e}$ corresponds to approximately $1.36 \times 10^{4}$ time steps for Cases 1–4 and $8.16 \times 10^{3}$ time steps for Cases 5–7. }
\label{fig:surface}
\end{figure}

The temporal evolution of the interfacial area is shown in Figure~\ref{fig:surface}. The approximately maintained interfacial area in all cases provides further evidence of statistical droplet-volume conservation. Within the phase-field framework, the interfacial area can be approximated using the coarea formula~\cite{federer1959curvature},
\begin{equation}
S \approx \frac{1}{\phi_{\rm D} - \phi_{\rm C}}\int_\Omega \lvert \boldsymbol{\nabla} \phi \rvert , dV.
\end{equation}
The interfacial area is reported in terms of its relative deviation from the initial value, $S(t)/S(0)-1$, where $S(0)=4\pi r_0^2$ and $r_0=(3\varphi_{\rm D}L^3/4\pi)^{1/3}$ denotes the initial droplet radius.

As expected, turbulence significantly stretches the interfaces in all cases, leading to a pronounced increase in interfacial area with increasing Weber number. The reduced surface tension at higher Weber numbers provides weaker resistance to droplet deformation and breakup. For Cases~4–7, which involve increasing density ratios, the degree of interfacial deformation decreases. This behavior can be attributed to two factors: first, higher droplet density increases inertia, enhancing resistance to deformation; second, the increased system inertia reduces overall turbulence intensity, thereby weakening the ability of turbulence to stretch the interfaces.

To further quantify droplet volume conservation, time-averaged droplet volumes and interfacial areas are summarized in Table~\ref{tab:averagedvolume}. The reported uncertainties correspond to a 95\% confidence interval, calculated as $1.96,\sigma_A/\sqrt{N}$, where $\sigma_A$ is the sample standard deviation. The effective number of independent samples is estimated as $N=T(0.5)/T_{\rm ave}$, with $T(0.5)$ denoting the time lag at which the autocorrelation decays to 0.5 and $T_{\rm ave}$ the total sampling duration. In all cases, the maximum deviation of the mean droplet volume from its initial value is less than 1\%, and this deviation can be further reduced by selecting a more optimal value of the tuning parameter $\alpha$.

\begin{table}[h]
    \centering
    \caption{Time-averaged droplet volume and interface area.}
    \label{tab:averagedvolume}
\resizebox{\textwidth}{!}{
\begin{tabular}{c c c c c c c c c c c}
        \hline
         &Case 1 & Case 2 & Case 3 & Case 4 & Case 5 & Case 6 & Case 7\\
        \hline
         $V_{\rm D}(t)/V_{\rm D}(0)$&$1.0003\pm0.0002$&$0.9980\pm0.0004$&$0.9930\pm0.0010$&$0.9912\pm0.0020$&$0.9977\pm0.0026$&$0.9989\pm0.0015$&$1.0023\pm0.0012$ \\
         \hline
         &Case 1&Case 2&Case 3&Case 4&Case 5& Case 6&Case 7\\
         \hline
$S(t)/S(0)-1$&$3.8294\pm0.0621$&$6.4756\pm0.0834$&$11.7322\pm0.1316$&$17.5325\pm0.1888$&$17.4642\pm0.2379$&$16.2515\pm0.1675$&$15.5882\pm0.1565$\\
        \hline
    \end{tabular} }    
\end{table}

A remaining concern is whether the modified CAC equation may induce artificial coarsening, namely unphysical mass transfer from smaller droplets to larger ones. While coarsening (Ostwald ripening) is a genuine physical process driven by free-energy minimization, it typically occurs on time scales much longer than those relevant to hydrodynamic evolution. In phase-field simulations, however, the presence of diffuse interfaces can significantly accelerate this process in an unphysical manner. In particular, nonlocal mass-correction strategies have been shown to promote artificial coarsening~\cite{chai2018comparative,hu2019hybrid}, because the mass lost through the dissolution of small droplets is redistributed to larger droplets~\cite{li2025comparativestudycriticalassessment}, leading to an artificial depletion of the small-droplet population. A similar behavior was observed for the global mass-correction model of Wang et al.~\cite{wang2015mass}, as discussed in Section~\ref{sec:previous}.

\begin{figure}

\begin{subfigure}[b]{0.5\textwidth}
        \centering
        \includegraphics[width=\textwidth]{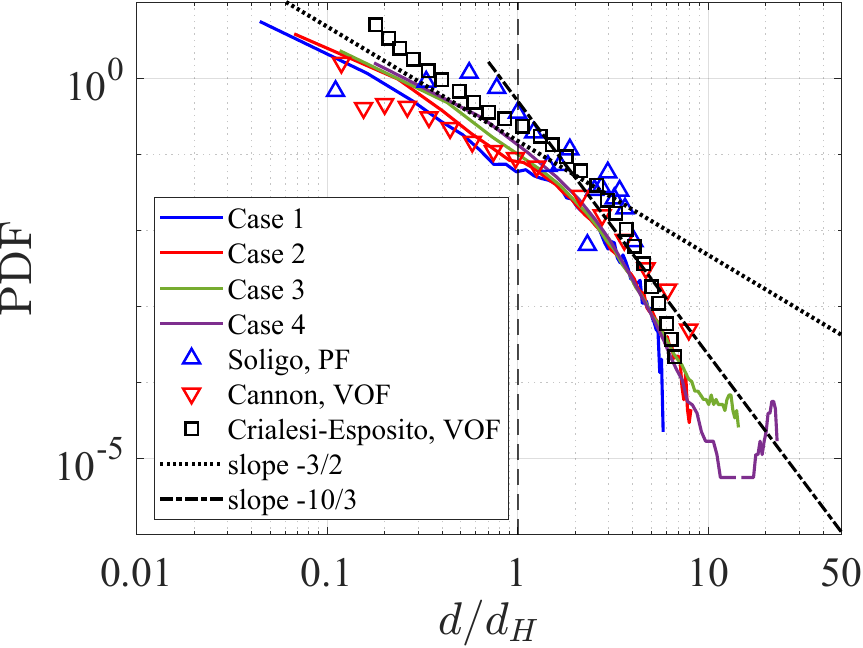}
        \caption{Case 1-4 with different We}
        \label{fig:dropletsize1}
\end{subfigure}
\hfill
    \begin{subfigure}[b]{0.5\textwidth}
        \centering
        \includegraphics[width=\textwidth]{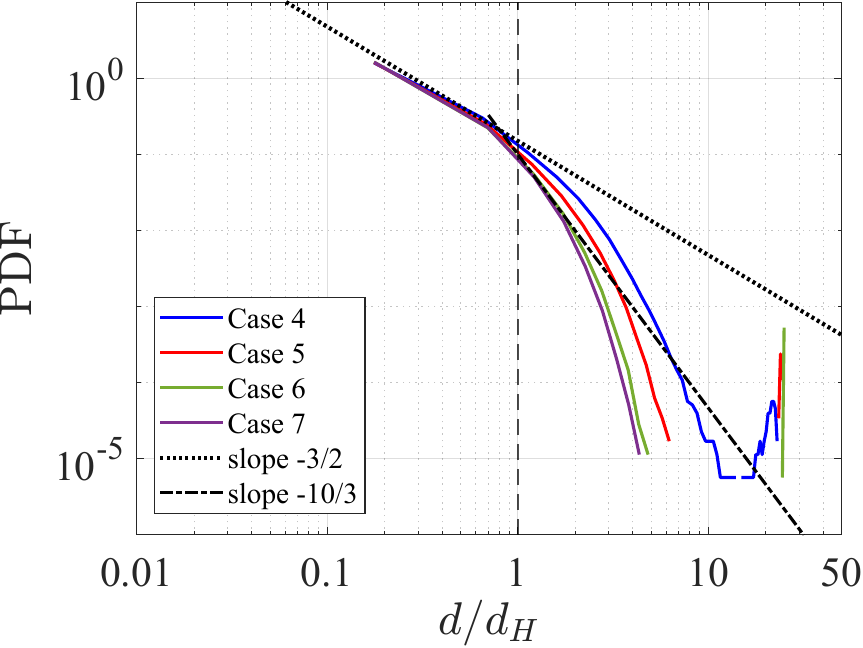}
        \caption{Case 4-7 with different density ratio}
        \label{fig:dropletsize2}
    \end{subfigure}

    \caption{Probability distribution functions (PDF) of the droplet size. Two power-law scalings are indicated: $d^{-3/2}$ in the coalescence-dominated regime and $d^{-10/3}$ in the breakage-dominated regime~\cite{rodriguez2025drop}. For comparison, data from phase-field (PF) simulations by Soligo et al.~\cite{soligo2019breakage}, volume-of-fluid (VOF) simulations by Crialesi-Esposito et al.~\cite{crialesi2022modulation}, and Cannon et al.~\cite{cannon2024morphology} are also shown.}
\label{fig:dropletsize}
\end{figure}

As already evident in Figures~\ref{fig:isosurfaces} and~\ref{fig:contours}, small droplets persist at late times when the proposed modified CAC equation is employed. This observation is further supported by the droplet size distributions shown in Figure~\ref{fig:dropletsize}, where each probability density function (PDF) is constructed from more than $10^{6}$ droplet samples. Droplet size distributions in HIT are known to exhibit two scaling regimes: for droplets smaller than the Hinze scale $d_{\rm H}$, coalescence dominates and the PDF follows a $d^{-3/2}$ scaling, whereas for droplets larger than $d_{\rm H}$, breakup dominates and the PDF decays as $d^{-10/3}$. Both scaling laws are well reproduced across different Weber numbers and density ratios. Importantly, the correct scaling behavior persists even in the small-droplet limit, indicating the absence of artificial coarsening.

For comparison, results from previous simulations of droplet-laden HIT~\cite{soligo2019breakage,crialesi2022modulation,cannon2024morphology} are also included. Since these reference studies were conducted at unity density ratio, comparisons are restricted to Cases~1–4. Volume-of-fluid (VOF) simulations, which do not suffer from droplet volume loss, agree well with theoretical predictions. In contrast, the phase-field simulations of Soligo et al.~\cite{soligo2019breakage} exhibit a pronounced deficit in the small-droplet regime, consistent with the dissolution effects reported in that work.

\begin{figure}
        \centerline{
        \includegraphics[width=0.6\textwidth]{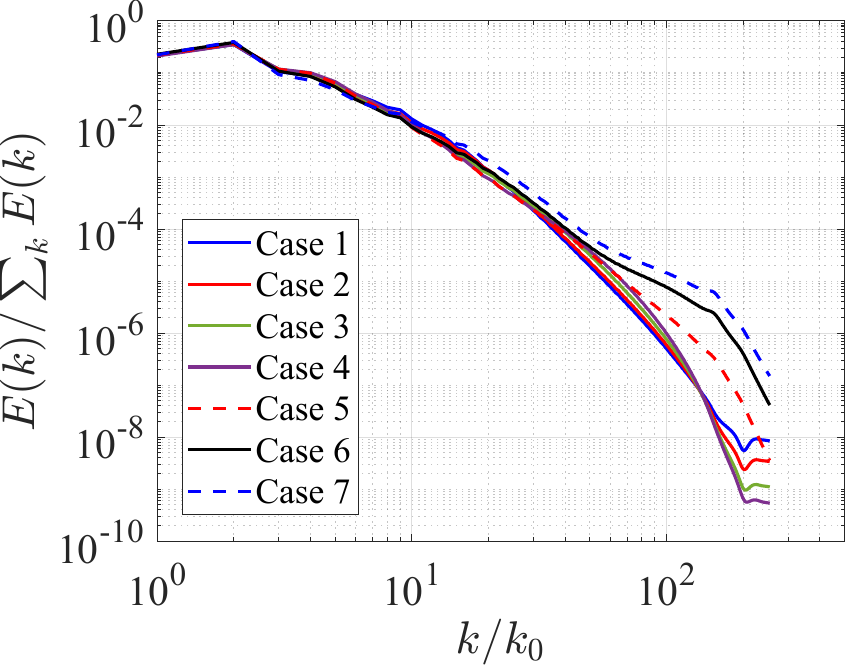}}
        \caption{Spectra of the turbulent kinetic energy}
        \label{fig:Energyspectra}
\end{figure}

Finally, phase-field simulations of droplet-laden turbulence are known to suffer from spurious currents—unphysical velocity fluctuations near interfaces that may contaminate small-scale turbulence statistics~\cite{komrakova2015numerical,elghobashi2019direct}. Figure~\ref{fig:Energyspectra} presents turbulent kinetic energy spectra for representative cases. A modest energy increase is observed at high wavenumbers, corresponding to the smallest resolved scales. However, its magnitude remains limited, particularly in cases with low surface tension and moderate density ratios. By comparison with VOF-based simulations~\cite{crialesi2022modulation} and with our previous studies of turbulence laden with finite-size particles—where similar high-wavenumber noise is observed near particle surfaces~\cite{peng2023parameterization}—we conclude that spurious currents in the present simulations are sufficiently weak and do not contaminate the physically relevant turbulence modulation induced by dispersed droplets.

\section{Conclusions}\label{sec:conclusion}

Accurate and robust droplet volume conservation remains a long-standing challenge in phase-field simulations of droplet-laden turbulent flows. In turbulent environments, continuous droplet deformation, breakup, and coalescence generate a wide spectrum of droplet sizes, including many droplets whose diameters are comparable to the interface thickness. These small droplets are particularly prone to excessive numerical dissolution, leading to systematic and cumulative droplet volume loss in long-time simulations. Existing remedies—including the CAC equation, modified CH formulations, singular mobility models, and global mass-correction strategies— have proven insufficient. They either fail to prevent volume loss under strong turbulence, suffer from numerical instability, or introduce unphysical side effects such as artificial coarsening and violation of mass conservation.

In this work, we propose a modified CAC formulation designed \emph{specifically} to achieve statistical droplet-volume conservation in droplet-laden turbulence. The key innovation is a curvature-dependent counter-diffusion term acting along the interface normal direction. This term selectively counteracts the excessive dissolution of small droplets with high curvature while remaining inactive for flat interfaces and sufficiently large droplets. Formulated in divergence form, this correction preserves global mass and momentum, avoids nonlocal redistribution of dissolved material, and consequently prevents  artificial coarsening. The model's parameterization is linked to Hinze’s law, ensuring it naturally reduces to the original formulation in cases where the original CAC is already sufficient. Importantly, the proposed correction can be incorporated directly into the existing CAC flux, imposing no additional computational or communication cost in parallel simulations.

Through direct numerical simulations of droplet-laden homogeneous isotropic turbulence over a wide range of Weber numbers and density ratios, we demonstrate that the proposed model successfully preserves droplet volume in a statistical sense over simulation times exceeding one million time steps—a regime where all previously proposed phase-field models examined here fail. The method maintains a physically realistic population of small droplets, reproduces Hinze-scale-based droplet size distributions, and avoids the spurious accumulation of mass in large droplets, a common artifact of global correction approaches. Analysis of the energy spectra further confirms that spurious currents remain weak and do not contaminate the physically relevant turbulent dynamics.

Overall, this work establishes a physically motivated and computationally efficient framework for conducting long-time, high-Weber-number phase-field simulations of droplet-laden turbulence. By shifting the focus from instantaneous volume enforcement to statistically consistent volume conservation, the proposed modified CAC equation resolves a fundamental limitation of prior phase-field models, enabling reliable DNS of multiphase turbulent flows over extended durations.

\section{Appendix A: The lattice Boltzmann solvers for various phase-field models}\label{sec:appendixA}
The LB method can be regarded as a specialized discretization framework for solving the hydrodynamic equations governing fluid motion, such as the phase-field equation and the Navier–Stokes equations. Rather than employing conventional finite-difference-type discretizations, the LB method solves these equations indirectly by evolving a set of particle distribution functions associated with discrete velocity sets. The macroscopic fields are then recovered from suitable moments of these distribution functions, which reproduce the target governing equations in the hydrodynamic limit.

In phase-field LB models, two distinct sets of distribution functions are typically employed: one for solving the phase-field equation and the other for the Navier–Stokes equations. For convenience of presentation, we recast the evolution rules of the distribution functions—commonly referred to as the lattice Boltzmann equations (LBE)—into their single-relaxation-time (BGK) form, even though numerical stabilization techniques such as the multi-relaxation-time (MRT) model are used in the actual simulations. The LBE read~\cite{yang2016lattice,liang2018phase}
\begin{subequations}
\begin{align}
& g_\alpha (\bm{x} + \bm{e}_\alpha \delta t, t + \delta t) 
= g_\alpha (\bm{x},t) 
- \frac{g_\alpha (\bm{x},t) - g_\alpha^{eq}(\bm{x},t)}{\tau_\phi} 
+ \left(1 - \frac{1}{2\tau_\phi}\right) \delta t \, G_\alpha (\bm{x},t),
\label{eq:LBEg} \\[2mm]
& f_\beta (\bm{x} + \bm{e}_\beta \delta t, t + \delta t) 
= f_\beta (\bm{x},t) 
- \frac{f_\beta (\bm{x},t) - f_\beta^{eq}(\bm{x},t)}{\tau} 
+ \left(1 - \frac{1}{2\tau}\right) \delta t \, F_\beta (\bm{x},t).
\label{eq:LBEf}
\end{align}
\label{eq:LBE}
\end{subequations}
Here, $\bm{x}$ and $t$ denote space and time, respectively. The distribution function $g_\alpha$ corresponds to the phase-field equation with discrete velocity ${\bm e}_\alpha$, while
${f_\beta }$ corresponds to the Navier–Stokes equations with discrete velocity ${\bm e}_\beta$. The velocity sets used for the phase-field and hydrodynamic solvers need not be identical. The relaxation times $\tau_\phi$ and $\tau$ are related to the local diffusivities via~\cite{yang2016lattice,liang2018phase}
\begin{equation}
\tau_\phi = \frac{M_\phi}{c_{s,g}^2} + \frac{1}{2},
\qquad \tau = \frac{\mu}{\rho c_{s,f}^2} + \frac{1}{2},
\label{eq:fakhari_tau_model}
\end{equation}
where $c_{s,g}$ and $c_{s,f}$ are the lattice sound speeds determined by the corresponding lattice velocity sets.

All phase-field models considered in this work share the same LBE structure; the differences lie only in the specific definitions of the equilibrium distribution functions $g_\alpha^{eq}$ and $f_\beta^{eq}$, as well as the source terms $G_\alpha$ and $F_\alpha$. 

\subsection{Locally corrected CH equations}

For the original CH equation and the three locally corrected variants (CH-PC, CH-FC, and CH-IC), the equilibrium distributions and source terms in the LBE are defined as follows:
\begin{subequations}
\begin{align}
&g_\alpha^{eq}(\bm{x},t) = 
\begin{cases}
\phi - (1 - w_\alpha) \mu_\phi + \phi \Gamma_\alpha({\bm u}), & \alpha = 0, \\[1pt]
w_\alpha  \mu_\phi +  \phi \Gamma_\alpha({\bm u}), & \alpha \ne 0,
\end{cases}
\label{eq:Yang_g_eq} \\[2mm]
&f_\beta^{eq}(\bm{x},t) = w_\beta p  + c_s^2  \rho \Gamma_\beta({\bm u}) .
\label{eq:Yang_f_eq}\\[2mm]
&G_\alpha(\bm{x},t) = - \frac{\phi}{c_{s}^2 \rho} (\bm{e}_\alpha - \bm{u}) \cdot (\boldsymbol{\nabla} p - \bm{F}) \,  \, s_\alpha(\mathbf{u}) - w_\alpha\frac{2}{\left(\tau_\phi - 1\right)\delta t}\frac{{\bm e}_\alpha\cdot {\bm S}}{c_s^2},
\label{eq:Yang_G} \\[2mm]
&F_\beta(\bm{x},t) = (\bm{e}_\beta - \bm{u}) \cdot \left[  \bm{F} \, s_\beta(\bm{u}) +  \Gamma_\beta(\bm{u}) \, c_s^2 \boldsymbol{\nabla} \rho \right] - \omega_\beta c_s^2 \rho \, \gamma \, \boldsymbol{\nabla} \cdot (M_\phi \boldsymbol{\nabla} \mu_\phi + {\bm S}),
\label{eq:Yang_F} \\[2mm]
\end{align}
\label{eq:Yang_eq}
\end{subequations}
where the auxiliary functions $s_\alpha$ ($s_\beta$) and $\Gamma_\alpha$ ($\Gamma_\beta$) are given by
\begin{equation}
    s_\alpha(\bm{u}) = w_\alpha\left[1 + \frac{\bm{e}_\alpha \cdot \bm{u}}{c_s^2} + \frac{(\bm{e}_\alpha \cdot \bm{u})^2}{2 c_s^4} - \frac{\bm{u} \cdot \bm{u}}{2 c_s^2}\right],\qquad \Gamma_\alpha(\bm{u}) = w_\alpha\left[\frac{\bm{e}_\alpha \cdot \bm{u}}{c_s^2} + \frac{(\bm{e}_\alpha \cdot \bm{u})^2}{2 c_s^4} - \frac{\bm{u} \cdot \bm{u}}{2 c_s^2}\right].
\end{equation}
Here, the lattice sound speed $c_s$ should be interpreted as $c_{s,g}$ or $c_{s,f}$, depending on whether it appears in the phase-field or hydrodynamic solver.

The total force ${\bm F}$ acting on the fluid consists of the surface-tension force ${\bm F}_{s} = \mu_\phi \boldsymbol{\nabla}\phi$ and the external body force ${\bm F}_{b} = \rho {\bm g}$~\cite{yang2016lattice}. The vector ${\bm S}$ represents the model-dependent correction flux associated with the phase-field equation and takes the following forms for the four cases considered:
\begin{subequations}
     \begin{align}
     {\rm Original~CH}:  \quad &{\bm S} = 0\\
     {\rm CH-PC:}\quad &{\bm S} = \lambda_{\rm PC}M_{\rm CH}\left[\boldsymbol{\nabla}\phi + \frac{4}{W}\frac{\left(\phi - \phi_{\rm D}\right)\left(\phi - \phi_{\rm C}\right)}{\phi_{\rm D} - \phi_{\rm C}}{\bm n}\right]\\
     {\rm CH-FC}:\quad & {\bm S} = \lambda_{\rm PC}M_{\rm CH}\left[\boldsymbol{\nabla}\phi + \frac{4}{W}\frac{\left(\phi - \phi_{\rm D}\right)\left(\phi - \phi_{\rm C}\right)}{\phi_{\rm D} - \phi_{\rm C}}{\bm n}\right] - M_{\rm CH}\left({\bm n}\cdot\boldsymbol{\nabla}\mu_{\phi}\right){\bm n}\\
     {\rm CH-IC}: \quad & {\bm S} = \lambda_{\rm IC}\vert \boldsymbol{\nabla}\left({\bm u}\cdot {\bm n}\right)\vert\left(\phi - \phi_{\rm D}\right)\left(\phi - \phi_{\rm C}\right){\bm n} 
    \end{align}   
\end{subequations}

The macroscopic hydrodynamic quantities are recovered from the distribution functions as
\begin{subequations}
    \label{eq:phi_u_p_update}
    \begin{align}
        \phi &= \sum_\alpha g_\alpha
        \label{eq:phi_update} \\ 
        {\bm u} &= \frac{1}{c_{s,f}^2\rho}\left(\sum_\beta f_\beta {\bm e}_\beta + \frac{1}{2}c_{s,f}^2{\bm F}\delta t\right)
        \label{eq:u_update} \\
        p &= \sum_{\beta}f_\beta + \frac{1}{2}\delta t c_{s,f}^2\Big[{\bm u}\cdot{\boldsymbol{\nabla}\rho}-\rho\frac{\rho_{\rm D} - \rho_{\rm C}}{\rho_{\rm C}\phi_{D} - \rho_{\rm D}\phi_{\rm C}}\boldsymbol{\nabla}\cdot\left(M_{\rm CH}\boldsymbol{\nabla}\mu_\phi + {\bm S}\right)\Big]
        \label{eq:p_update}
    \end{align}
\end{subequations}

The local density and dynamic viscosity are interpolated linearly between the bulk-phase values according to the local order parameter,
\begin{equation}
    \rho = \rho_{\rm C} + \frac{\phi - \phi_{\rm C}}{\phi_{\rm D} - \phi_{\rm C}}\left(\rho_{\rm D} - \rho_{\rm C}\right),\quad \mu = \mu_{\rm C} + \frac{\phi - \phi_{\rm C}}{\phi_{\rm D} - \phi_{\rm C}}\left(\mu_{\rm D} - \mu_{\rm C}\right).
\end{equation}
When ${\bm S} = 0$, the above formulation reduces to the original phase-field LB model proposed by Yang et al~\cite{yang2016lattice}.

\subsection{Modified CH equation with singular mobility}
In principle, the two modified CH models with singular mobility can be implemented by directly replacing the constant mobility in the original CH equation with a spatially varying one. However, when the mobility approaches zero in the bulk phases, the local Peclet number becomes unbounded, which renders the standard phase-field LB solver numerically unstable. This difficulty is well known in LB implementations of degenerate or singular diffusion operators.

To overcome this issue, a lattice kinetic scheme is adopted~\cite{inamuro2002lattice}. The key idea is to introduce gradient-dependent terms directly into the equilibrium distribution functions, thereby relaxing the strict coupling between the physical mobility and the relaxation time. This treatment significantly enhances numerical stability, especially in regions where the mobility is very small. For the two singular-mobility models considered here—namely, the models proposed by Bao and Guo~\cite{bao2024phase}, and by Yang and Kim~\cite{yang2025two}—the equilibrium distribution functions for both the phase-field and hydrodynamic solvers are modified as
\begin{subequations}
\begin{align}
g_\alpha^{eq}(\bm{x},t) &=
\begin{cases}
\phi - (1-w_\alpha) \mu_\phi + \phi \, {\Gamma}_\alpha(\bm{u}) + w_\alpha A \delta t  \bm{e}_\alpha \cdot \boldsymbol{\nabla} \mu_\phi, & \alpha = 0, \\[1pt]
w_\alpha  \mu_\phi +  \phi \, {\Gamma}_\alpha(\bm{u}) + w_\alpha A \delta t \bm{e}_\alpha \cdot \boldsymbol{\nabla} \mu_\phi, & \alpha \ne 0,
\end{cases}
\label{eq:bao_g_eq} \\[2mm]
f_\beta^{eq}(\bm{x},t) &= w_\beta p + c_s^2 \rho \, {\Gamma}_\beta(\bm{u})  
+ \frac{1}{2} \rho w_\beta B \delta t \, \mathbf{S} : \left( \bm{e}_\beta \bm{e}_\beta - c_s^2 \mathbf{I} \right),
\label{eq:bao_f_eq} 
\end{align}
\label{eq:bao_eq}
\end{subequations}
where $\mathbf{S} = \boldsymbol{\nabla}\bm{u} + (\boldsymbol{\nabla}\bm{u})^T$ denotes the strain-rate tensor. This tensor should be distinguished from the correction flux ${\bm S}$ introduced in the locally corrected CH models. Here, $\mathbf{I}$ is the identity tensor. The parameters $A$ and $B$ are auxiliary free parameters that locally decouple the physical mobility and kinematic viscosity from the corresponding relaxation times. Specifically, the mobility $M_\phi$ and the kinematic viscosity $\nu$ are related to the relaxation times through
\begin{subequations}
\begin{align}
M_\phi &= c_{s,g}^2  \delta t \left(\tau_\phi -\frac{1}{2} - A\right), 
\label{eq:bao_Mphi} \\[2mm]
\nu &= c_{s,f}^2 \delta t \left(\tau -\frac{1}{2} - B\right),
\label{eq:bao_v}
\end{align}
\label{eq:bao_tau}
\end{subequations}
respectively.

In practical simulations, the relaxation times $\tau_\phi$ and $\tau$ are kept constant to maintain numerical stability, while the parameters $A$ and $B$ are adjusted locally to reproduce the prescribed spatial variations of the mobility and viscosity. The source terms $G_\alpha$ and $F_\beta$, as well as the procedures for updating the macroscopic hydrodynamic quantities, remain unchanged and follow the corresponding definitions used in the LB model for the original CH equation. 

\subsection{Modified CH equation with global mass constraint}
For the modified CH model with global mass correction proposed by Wang et al., the LB implementation differs slightly from that of the other models considered in this work, because the correction term relies on global information and therefore cannot be incorporated into the diffusion flux of the original CH equation.

In this model, the equilibrium distribution functions for both the phase-field solver and the hydrodynamic solver, as well as the source term in the hydrodynamic solver, are identical to those given in Eqs.~\eqref{eq:Yang_g_eq} and \eqref{eq:Yang_f_eq}.
The only modification appears in the source term of the phase-field solver, where an additional contribution associated with the global mass-correction term $Q_G$ is introduced. Specifically, the source term reads
\begin{equation}
G_\alpha(\bm{x},t)
=
- \frac{\phi}{c_{s,g}^2 \rho}
(\bm{e}_\alpha - \bm{u}) \cdot (\boldsymbol{\nabla} p - \bm{F})
\, s_\alpha(\mathbf{u})
+ w_\alpha Q_{\rm G} .
\end{equation}

As a consequence, the macroscopic order parameter must be updated as
\begin{equation}
\phi = \sum_\alpha g_\alpha + \frac{1}{2} Q_{\rm G}\delta t .
\end{equation}
The velocity and pressure are updated in the same manner as in Eqs.~\eqref{eq:u_update} and~\eqref{eq:p_update}, with $\bm{S}=0$.

\subsection{Conservative AC equation}

The LB formulation for the CAC equation differs from that used for CH-type models, owing to the different mathematical structures of the governing equations. In particular, the CAC equation does not involve a fourth-order diffusion operator, which allows for a more straightforward and numerically robust LB implementation. As several well-established phase-field LB models for the CAC equation are already available in the literature, these formulations can be adopted directly without further modification.

In the present work, we employ the LB model proposed by Liang et al.~\cite{liang2018phase}, which has been extensively validated for two-phase flow simulations involving complex interfacial dynamics. In this model, the equilibrium distribution functions and the corresponding source terms for the phase-field and hydrodynamic solvers are defined as~\cite{liang2018phase}
\begin{subequations}
\begin{align} 
g_\alpha^{eq}(\bm{x},t) &= w_\alpha \, \phi \left( 1 + \frac{\bm{e}_\alpha \cdot \bm{u}}{c_{s,g}^2} \right),
\label{eq:liang_geq} \\[2mm]
f_\beta^{eq}(\bm{x},t) &=
\begin{cases}
\dfrac{p}{c_{s,f}^2} (w_\beta - 1) + \rho  \Gamma_{\beta}\left({\bm u}\right) 
, & \beta = 0, \\[1mm]
\dfrac{p}{c_{s,f}^2} w_\beta + \rho \Gamma_{\beta}\left({\bm u}\right), & \beta \neq 0,
\end{cases}
\label{eq:liang_feq}\\[2mm]
G_\alpha(\bm{x},t) 
&= \omega_\alpha \, \bm{e}_\alpha \cdot \frac{\partial_t (\phi \bm{u})}{c_{s,g}^2} 
- \omega_\alpha 
 \frac{4}{W}\frac{\left(\phi - \phi_{\rm D}\right)\left(\phi - \phi_{\rm C}\right)}{\phi_{\rm D} - \phi_{\rm C}} \left(\bm{e}_\alpha \cdot {\bm n}\right),
\label{eq:liang_G_final} \\[2mm]
F_\beta(\bm{x},t) &= \omega_\beta \left[ \bm{u} \cdot \boldsymbol{\nabla} \rho + \frac{\bm{e}_\beta \cdot \bm{F}}{c_{s,f}^2} + \frac{\bm{u}\boldsymbol{\nabla} \rho : (\bm{e}_\beta \bm{e}_\beta - c_{s,f}^2 \mathbf{I})}{c_{s,f}^2} \right],
\label{eq:liang_F_final}
\end{align}
\label{eq:liang_eqs}
\end{subequations} 
where the second term in $G_\alpha$ represents the conservative interface-compression flux that counteracts numerical diffusion of the order parameter normal to the interface.

The macroscopic hydrodynamic variables are recovered from the distribution functions as~\cite{liang2018phase}
\begin{equation}
\phi = \sum_{\alpha} g_\alpha, \quad  \bm{u} = \frac{1}{\rho}\left[\sum_\beta \bm{e}_\beta f_\beta + \frac{1}{2} \bm{F}\delta t\right],  
 \quad p = \frac{c_{s,f}^2}{1-\omega_0} \left[ \sum_{\beta \neq 0} f_\beta + \frac{\delta t}{2} \bm{u} \cdot \boldsymbol{\nabla} \rho 
+ \rho  w_0 \left( - \frac{\bm{u} \cdot \bm{u}}{2 c_{s,f}^2} \right) \right]
\end{equation}

\subsection{The proposed modified CAC equation}
In the proposed modified CAC model, the additional correction term is constructed such that it can be naturally incorporated into the original CAC equation without altering its fundamental structure. As a result, the LB realization of the modified CAC model remains almost identical to that of the baseline CAC model described in the previous subsection. In practice, this modification affects only the definition of the source term associated with the phase-field equation, while all other components of the LB solver—including the equilibrium distribution functions, the hydrodynamic solver, and the macroscopic variable updates—remain unchanged.

Specifically, the source term $G_\alpha$ in the phase-field LB equation is modified as
\begin{equation}
    G_\alpha(\bm{x},t) = \omega_\alpha \, \bm{e}_\alpha \cdot \frac{\partial_t (\phi \bm{u})}{c_{s,g}^2} 
- \omega_\alpha
 \left[\frac{4}{W}\frac{\left(\phi - \phi_{\rm D}\right)\left(\phi - \phi_{\rm C}\right)}{\phi_{\rm D} - \phi_{\rm C}} - a\right]\left(\bm{e}_\alpha \cdot {\bm n}\right),
\end{equation}
where the newly introduced parameter $a$ controls the strength of the additional correction.

\section{Appendix B: Estimation of the maximum droplet-volume variation in phase-field simulations of droplet-laden turbulence}\label{sec:appendixB}

As argued repeatedly in the main text, it is not reasonable to enforce strict droplet-volume conservation in phase-field simulations of droplet-laden turbulence. The fundamental reason is that the use of a diffuse interface inevitably leads to apparent droplet-volume “non-conservation” when simple threshold-based criteria—such as identifying the droplet region by $\phi \ge \phi_0$—are employed. In this appendix, we provide a mathematical justification for this statement and offer a rough estimate of the magnitude of droplet-volume fluctuations that may arise even when the underlying phase-field model conserves the order parameter exactly. This analysis also serves to assess whether the volume fluctuations observed with the modified CAC equation are physically and numerically reasonable.

In standard phase-field models, the strictly conserved quantity is the integral of the order parameter,
\begin{equation}
    M = \int_\Omega \phi dV.
\end{equation}
However, when droplets deform, stretch, and break up under turbulent forcing, the droplet volume measured using a simple criterion such as $\phi \ge 0.5$ (here we assume $\phi \in [0,1]$ for simplicity) is generally not conserved.

The key issue is that the criterion $\phi \ge 0.5$ systematically excludes part of the diffuse interface. Let $S$ denote the total interfacial area and $W$ the interface thickness. The effective volume bias induced by the diffuse interface then scales as
\begin{equation}
    V_{\phi\ge 0.5} = V_{\rm 0} - CWS
\end{equation}
where $V_0 = \frac{4}{3}\pi r_0^3$ is the true liquid volume of a spherical droplet of radius $r_0$, and $C = \mathcal{O}(1)$ is a coefficient that depends weakly on the detailed profile of $\phi$ across the interface.

Consider first an initial state containing a single spherical droplet, for which 
\begin{equation}
    V_{\phi\ge0.5}^{0} = V_0 - CW\left(4\pi r_0^2\right).
\end{equation}
Under turbulent breakup, suppose that this droplet fragments into $N$ equal-sized droplets. The total interfacial area then increases by a factor
\begin{equation}
    \frac{S_N}{S_{0}} = \frac{N\cdot 4\pi r_{\rm small}^2}{4\pi r_0^2} = N^{1/3},
\end{equation}
where $r_{\mathrm{small}} = r_0 N^{-1/3}$. The corresponding $\phi \ge 0.5$ volume becomes 
\begin{equation}
    V_{\phi\ge0.5}^{N} = V_0 - CW\left(4\pi r_0^2 N^{1/3}\right).
\end{equation}
The apparent change in droplet volume is therefore
\begin{equation}
    \Delta V = V_{\phi\ge0.5}^{N} - V_{\phi\ge0.5}^{0} = - CW\left(4\pi r_0^2\right)\left(N^{1/3} - 1\right).
\end{equation}
Normalizing by the true droplet volume $V_0$ yields
\begin{equation}
    \frac{\vert\Delta V\vert}{V_0} = 3C\frac{W}{r_0}\left(N^{1/3} - 1\right). 
\end{equation}

We now estimate the coefficient $C$. Consider the standard one-dimensional equilibrium interface profile ($x\in(-\infty,\infty)$),
\begin{equation}
    \phi(x) = \frac{1}{2}\left[1+\tanh\left(\frac{x}{W}\right)\right].
\end{equation}
Using the criterion $\phi \ge 0.5$ to identify the liquid phase, the volume bias per unit interfacial area may be defined as
\begin{equation}
    \delta v \equiv \frac{\Delta V}{\Delta S} = \int_{0}^{+\infty}\left[1 - \phi(x)\right]dx = \frac{W\ln 2}{2},
\end{equation}
Since this contribution accounts for only one side of the interface, the total volume bias per unit area is $W \ln 2$. Recalling the definition
\begin{equation}
    \Delta V = -CW\Delta S,
\end{equation}
we immediately obtain $C = \ln 2\approx 0.693$. Although the exact equilibrium profile for a curved interface is generally unavailable, the hyperbolic tangent profile provides a reasonable estimate for the present purpose.

Substituting this value of $C$ gives
\begin{equation}
    \frac{\vert\Delta V\vert}{V_0} = 2.079\frac{W}{r_0}\left(N^{1/3} - 1\right).
\end{equation}
For the simulations reported in Section~\ref{sec:validation}, where $W = 3\delta x$, $r_0 = 147.4\delta x$, and where the interfacial-area amplification factor lies in the range $S_N / S_0 \approx 4.8$–$18.5$ ((see Table~\ref{tab:averagedvolume}), the resulting apparent volume deviation can readily exceed $10\%$. For the simulations discussed in Section~\ref{sec:previous}, which employ a wider interface thickness $W = 6\delta x$, and a smaller droplet radius $r_0 = 30\delta x$, the temporal fluctuations of the droplet volume estimated using the $\phi \ge0.5$ criterion are expected to be even more pronounced.

It should be emphasized that this estimate likely represents an upper bound, as it implicitly assumes that the entire interface contributes uniformly to the volume bias. Nevertheless, this analysis clearly demonstrates that enforcing strict droplet-volume conservation based on thresholded phase-field variables constitutes an overconstraint unless the liquid region can be identified with high accuracy. It also provides a quantitative explanation for the seemingly large volume fluctuations observed in turbulent phase-field simulations and supports the rationale for pursuing statistical, rather than pointwise, volume conservation in droplet-laden turbulence.

\textbf{Statement for Data Availability}: The source codes (written in FORTRAN 90) and the data for selected test cases reported in this work are available from the corresponding author upon reasonable request.

{\bf Acknowledgements}: This work has been supported by the National Natural Science Foundation of China 
(grant number 12472254, U2241269),  the Natural Science Foundation of Shandong Province, China (grant number ZR2024QA083), and the Fundamental Research Program of Shanxi Province(grant number 202303021211160).

\bibliography{mybibfile}

\end{document}